\documentclass{pinchcr}   

\usepackage{amsmath,amssymb,amsfonts}
\usepackage{graphicx}

\def\EF{E_{\text{F}}}
\def\kF{k_{\text{F}}}\def\uF{v_{\text{F}}}
\def\kB{k_{\text{B}}}

\def\npart{\cN}     
\def\partdensity{n} 
\def\dos{\cN}       
\def\dosv{n}        

\def\cN{\mathcal{N}}

\def\vk{\vec{k}}

\def\vr{\vec{r}}

\def\vS{\vec{S}}

\def\openone{\leavevmode\hbox{\small1\kern-3.3pt\normalsize1}}
\def\down{\downarrow}\def\up{\uparrow}

\renewcommand{\Im}{\operatorname{Im}}
\renewcommand{\Re}{\operatorname{Re}}

\newcommand{\step}{\operatorname{\theta}}
\def\Fermi{f_{\text{F}}}

\newcommand{\cref}[1]{Chapter~\ref{#1}}
\newcommand{\sref}[1]{Section~\ref{#1}}

\newcommand{\fref}[1]{Fig.~\ref{#1}}

\newcommand{\acomm}[2]{[#1,#2]_+}

\newcommand{\ket}[1]{\left\vert#1\right\rangle}

\newcommand{\bra}[1]{\langle#1\vert}

\newcommand{\hc}[1]{#1^{\dagger}}

\def\eps{\varepsilon}
\def\void{\ket{\varnothing}}
\def\Fermisea{\ket{\text{F}}}

\title{Interactions in quantum fluids}

\author{T. Giamarchi}
\affiliation{DPMC-MaNEP, University of Geneva, 24 Quai Ernest Ansermet, 1211 Geneva 4, Switzerland}
\authors{1}

\begin{document}

\maketitle

%
%
%
%
%
\tableofcontents

\maintext

\part{First part}

First part

\chapter{Interactions in quantum fluids}

\section{Introduction}

The problem of interacting quantum particles is one of the most
fascinating problems in physics, with an history nearly as long
as quantum mechanics itself. From a pure fundamental point of
view this is a staggering problem. Even in a one gram of solid
there are more particles interacting together than there are
stars in the universe. In addition these particles behave as
waves, since quantum effects are important and must obey
symmetrization or anti-symmetrization principles. It it thus no
wonder that despite nearly one century of efforts we still lack
the tools to give a complete solution of this problem, and that
sometimes even the concepts needed to describe such interacting
systems need to be sharpened.

The problem is not simply an academic one however. Indeed with
the discovery of quantum mechanics, came the understanding of
the behavior of electrons in solids and the band structure
theory. One of the fantastic success of such a theory was to
understand why some materials are metals, insulators or
semiconductors, an understanding which is at the root of the
discovery of the transistor and all the modern electronic
industry. An additional piece of knowledge was added by Landau,
with the so called Fermi liquid theory, where he showed that in
most fermionic systems the effects of interactions could be
essentially forgotten, and hidden in the redefinition of simple
parameters such as the mass of the particles. Being able to
forget about interactions paved the way to study the effects on
the real electronic systems of much smaller perturbations than
the electron-electron interactions, for example the electron
and lattice vibrations. This is at the root of the
understanding of many possible orders of the solids, such as
superconductivity and magnetism, or the interplay of magnetism
and conducting properties of the systems such as the giant
magnetoresistance. These properties have also had a profound
impact on our everyday's life.

Now such systems have been intensively studied and the
forefront of research has moved to materials for which we
cannot hide the effects of interactions anymore. This include
in particular the wide class of materials such as the oxides
that have properties as varied as being the best
superconductors known to date, or exhibit properties such as
ferroelectricity. These compounds or others that remain to be
discovered are clearly those who will be the materials of the
future for applications, and mastering of their properties
requires the deep understanding of the effects of interactions.
In addition to problems in condensed matter, cold atomic gases
have provided recently for marvellous realizations of such
strongly correlated systems and have thus added both to the
challenge we have to face, but also provided model systems that
could help us to make progress in that difficult field.

In these notes I review the basic concepts of the effects of
interactions on quantum particles. I focuss here mostly on the
case of fermions, but several aspects of interacting bosons
will be mentioned as well. The course has been voluntarily kept
at an elementary level and should be suitable for students
wanting to enter this field. I review the concept of Fermi
liquid, and then move to a description of the interaction
effects, as well as the main models that are used to tackle
these questions. Finally I study the case of one dimensional
interacting particles that constitutes a fascinating special
case.

\section{Fermi liquids}

This section is based on a master course given at the
university of Geneva, over several years together with C.
Berthod, A. Iucci, P. Chudzinski. Many more details can be
found on the course notes on
\verb+http://dpmc.unige.ch/gr_giamarchi/+.

\subsection{Weakly interacting fermions}

Let me start by recalling briefly some well known but important
facts about non interacting fermions. I will not recall the
calculation since they can be found in every textbook on solid
state physics \cite{ashcroft_mermin_book,ziman_solid_book}, but
just give the main results. These will be important for the
case of interacting electrons.

\begin{figure}
\begin{center}
\includegraphics[width=0.9\linewidth]{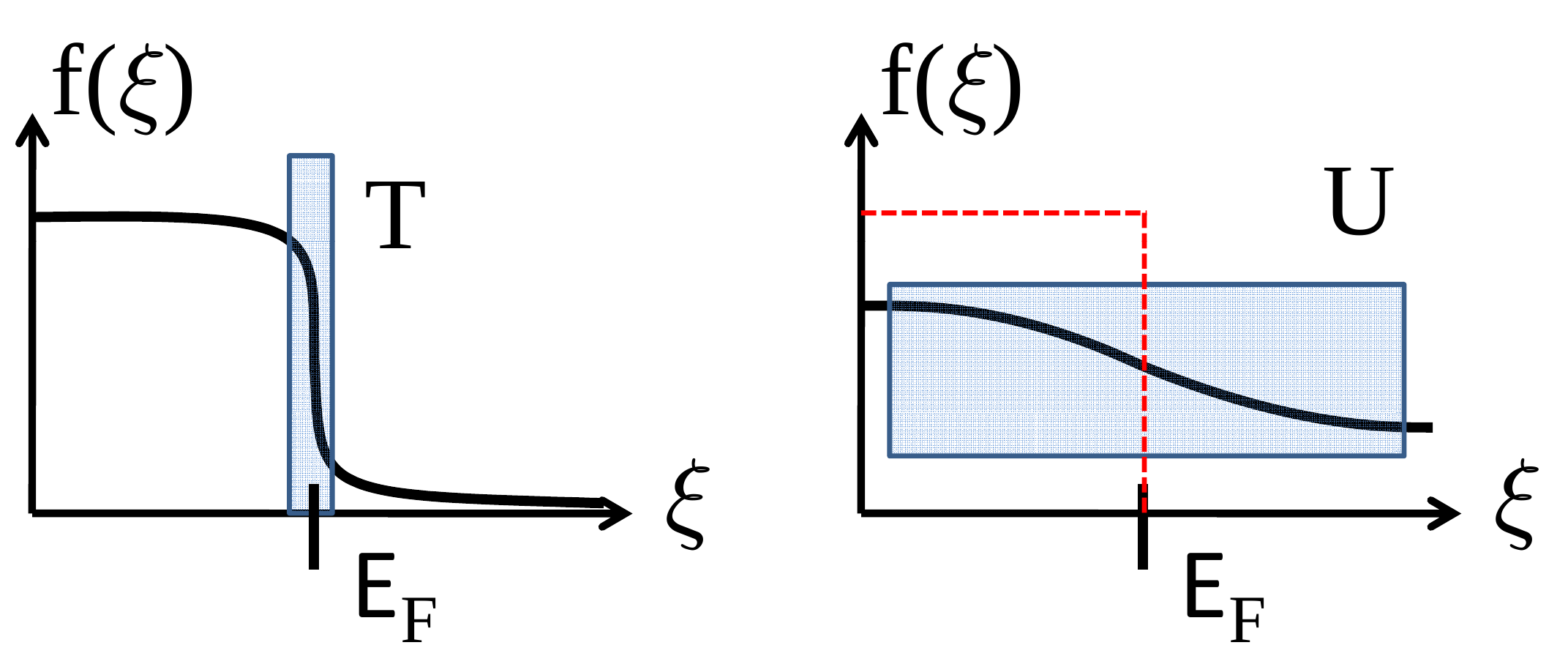}
\end{center}
\caption{\label{fig:broadenings} (left) Broadening of the Fermi distribution due to the temperature $T$.
Because in most solids the temperature $T$ is much smaller than the Fermi energy $\EF$ only a tiny fraction
close to the Fermi level can be excited. These fermions control all the properties of the system. (right) If the
interaction $U$ was acting as a temperature and was broadening the Fermi distribution one would get an extremely
large broadening. All the consequences of the sharp Fermi distribution (specific heat linear in temperature, compressibility
and spin susceptibility going to a constant at low temperatures) would be lost.}
\end{figure}

We consider independent electrons described by the Hamiltonian
\begin{equation} \label{eq:kinetic3}
 H_{\rm kin} = \sum_{\vk\sigma} \eps(\vk) \hc{c}_{\vk\sigma}c_{\vk\sigma}
\end{equation}
where the sum over $\vk$ runs in general over the first
Brillouin zone. One usually incorporates the chemical potential
in the energy $\xi(\vk)= \eps(\vk) - \EF$ to make sure that
$\xi(\vk) = 0$ at the Fermi level. The ground state of such
system is the unpolarized Fermi sea
\begin{equation} \label{eq:fermi_sea}
 \Fermisea = \prod_{\vk,\xi(\vk) \leq 0} \hc{c}_{\vk\up} \hc{c}_{k\down} \void
\end{equation}
At finite temperature states are occupied with a probability
\begin{equation}
 n(k) = \langle \hc{c}_{\vk} c_{\vk} \rangle = \Fermi(\xi(\vk)) = \frac1{e^{\beta\xi(\vk)}+1}
\end{equation}
given by the Fermi factor. A very important point, true for
most solids, is that the order of magnitude of the Fermi energy
is $\EF \sim 1 eV \sim 12000 K$. This means that the
temperature, or most of the energies that are relevant for a
solid (for example $30 GHz \sim 1K$) are extremely small
compared to the Fermi energy. As a result the broadening of the
Fermi distribution is extremely small. The important states are
thus be the ones in a tiny shell close to the Fermi level, as
shown in \fref{fig:broadenings}.
The other excitations are completely blocked by the Pauli
principle. This hierarchy of energies is of course what confers
to fermions in solids their unique properties and make them so
different from a classical system. As a consequences some of
the response of such a fermion gas are rather unique. The
specific heat is linear with temperature (contrarily to the
case of a classical gas for which it would be a constant)
\begin{equation}
 C_V(T) \propto \kB^2 \dos(\EF) T
\end{equation}
where $\dos(\EF)$ is the density of states at the Fermi level,
and $\kB$ the Boltzmann constant. The compressibility of the
fermion gas goes to a constant in the limit $T \to 0$, and the
same goes for the spin susceptibility, namely the magnetization
$M$ of the electron gas in response to an applied magnetic
field $H$
\begin{equation}
 \chi = \left(\frac{d M}{d H}\right)_T
\end{equation}
Note that a system made of independent spins would have had a
divergent spin susceptibility when $T\to 0$ instead of a
constant one. The slope of the specific heat, the
compressibility and the spin susceptibility of the free fermion
gas are all controlled, by the same quantity namely the density
of states at the Fermi level.

One could thus wonder what would be the effects of interactions
on such a behavior. Because of the interactions, the energy of
a particle can now fluctuate since the particle can give of
take energy from the others. Thus one could naively imagine
that the interactions produce an effect on the distribution
function similar to the one of a thermal bath, with an
effective ``temperature'' of the order of the strength of the
interaction. In order to determine the consequences of such a
broadening, one needs to estimate the strength of the
interactions. In a solid the interaction is mostly the Coulomb
interaction. However, in a metal this interaction is screened
beyond a length $\lambda$ that one can easily compute in the
Thomas-Fermi approximation
\cite{ashcroft_mermin_book,ziman_solid_book}
\begin{equation}
\lambda^{-2} = \frac{e^2 \dos(\EF)}{\epsilon_0}
\end{equation}
where $e$ is the charge of the particles and $\epsilon_0$ the
dielectric constant of the vacuum. To estimate $\lambda$ we can
use the fine structure constant
\begin{equation} \label{eq:fine_structure}
 \alpha =\frac{e^{2}}{4 \pi \epsilon_{0} \hbar c} =\frac{1}{137}
\end{equation}
to obtain
\begin{equation}
 \lambda^{-2} = 4\pi \alpha \hbar c \dos(\EF) = 4\pi \alpha \hbar c \frac{3n}{2 \EF}
\end{equation}
where $n$ is the density of particles. Using the density of
states per unit volume of free fermions in three dimensions
\begin{equation} \label{eq:density_states}
 \dosv(\eps) = \frac{m}{2 \pi^2 \hbar^2}\left(\frac{2 m \EF}{\hbar^2}\right)^{1/2} =  \frac32 \frac{\partdensity}{\EF}
\end{equation}
and $\EF = \hbar \uF \kF$, and $6 \pi^2 n = \kF^3$ one gets
\begin{equation}
\lambda^{-2} = \frac{1}{\pi} \alpha \frac{c}{\uF} \kF^2
\end{equation}
Since $c/\uF \sim 10^{2}$ in most systems, one finds that $\kF
\lambda \sim 1$. The screening length is of the order of the
inverse Fermi length, i.e. essentially the lattice spacing in
normal metals. This is a striking result: not only the Coulomb
interaction is screened, but the screening is so efficient that
the interaction is practically \emph{local}. We will use
extensively this fact in the definition of models below. Let us
now estimate the order of magnitude of this screened
interaction. The interaction between two particles can be
written as
\begin{equation}
 H_{\text{int}}= \int d\vr d\vr' V(\vr-\vr') \rho(\vr) \rho(\vr')
\end{equation}
Since the interaction is screened it is convenient to replace
it by a local interaction. Given our previous result let us
simply replace the screening length by $a$ the fermion-fermion
distance. The effective potential seen at point $\vr$ by one
particle is
\begin{equation}
 \int d\vr' V(\vr-\vr') \rho(\vr')
\end{equation}
Due to screening we should only integrate within a radius $a$
around the point $r$. Assuming that the density is roughly
constant one obtains
\begin{equation}
 \int_{|\vr-\vr'|<a} d\vr \frac{e^2}{4\pi \epsilon_0 |\vr-\vr'|} \rho_0 \sim  \frac{e^2\rho_0 S_d a^{d-1}}{4(d-1)\pi \epsilon_0}
\end{equation}
where $S_d$ is the surface of the sphere in $d$ dimensions.
Using $\rho_0 \sim 1/a^d$ and (\ref{eq:fine_structure}) one
gets
\begin{equation}
  \frac{S_d \alpha \hbar c}{(d-1) a}
\end{equation}
This potential acting on a particle has to be compared with the
kinetic energy of this particle at the Fermi level which is
$\EF = \hbar \uF \kF$. Since $\kF \sim a^{-1}$ one has again to
compare $\alpha$ and $c/\uF$. The two are about the same order
of magnitude. The Coulomb energy, even if screened (i.e. even
in a very good metal), is thus of the \emph{same} order of
magnitude than the kinetic energy. This means, for solids,
typical energies of the order of the electron volt. If such an
interaction was acting as a temperature in smearing the Fermi
function this would lead to an enormous smearing as shown in
\fref{fig:broadenings}.

This would be in complete contradiction with data on most of
solids. The specific heat in real materials is found to be
linear, albeit with a slope different from the naive free
electron picture at temperatures much smaller than the scale of
the interactions (see e..g Fig.~1.8 in
\cite{ashcroft_mermin_book}). This would be totally impossible
if the interactions had smeared the Fermi distribution to a
practically flat distribution. Similarly spin susceptibility
and compressibility are still found, e.g. in $^3$He to be
essentially constant at low
temperature\cite{greywall_landau_1,greywall_landau_2}, again
implying that the Fermi distribution must remain quite sharp.

\begin{figure}
\begin{center}
\includegraphics[width=0.8\linewidth]{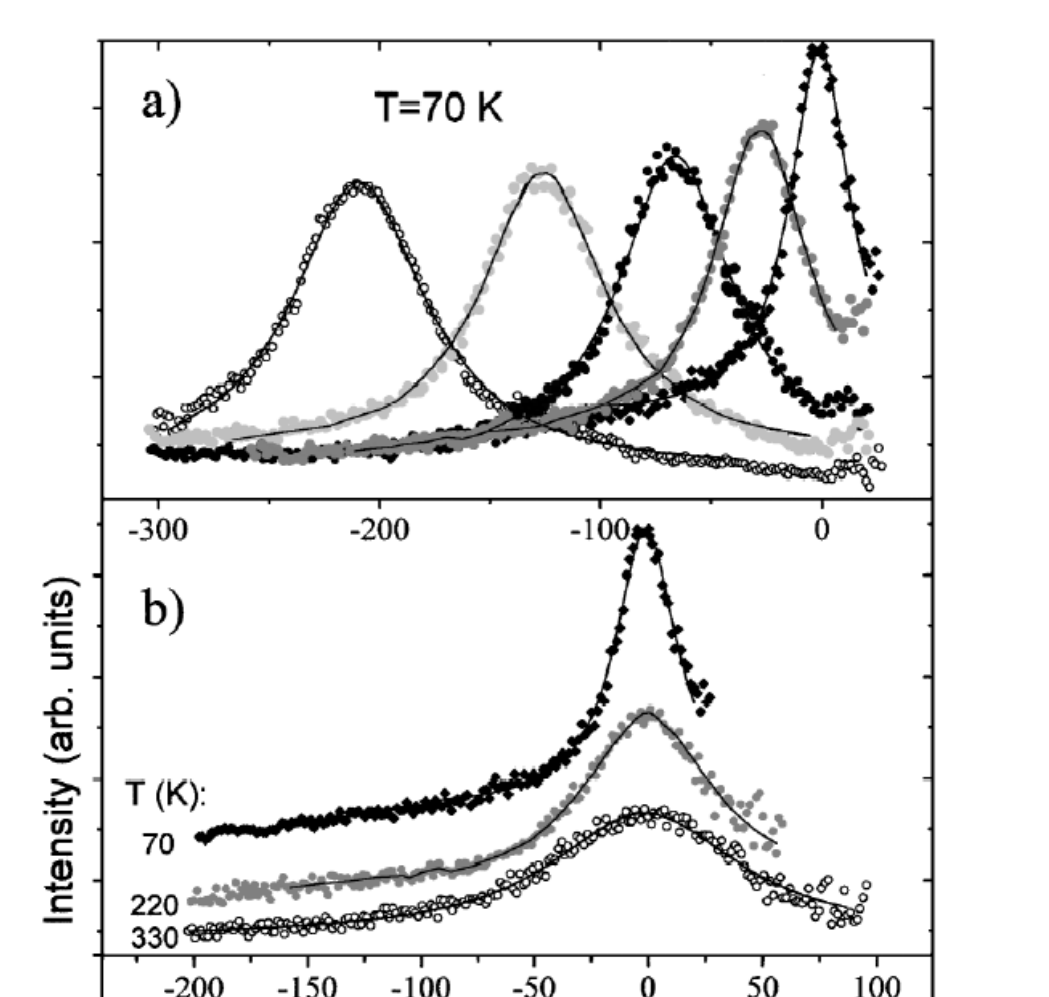}
\end{center}
\caption{\label{fig:photo_exp} Photoemission data from a Mo (110) surface.
(Top) The spectral function $A(\vk,\omega)$ is plotted
as a function of $\omega$. The zero denotes the Fermi level. Different peaks corresponds to different values of $\vk$.
One sees that, contrarily to naive expectations, the peaks in the spectral function
become narrower and narrower as one gets closer to the Fermi energy.
(Bot) Width of a peak close to the Fermi level as a function of the temperature $T$.
One sees that the width of the peak is controlled in a large part by the temperature,
which corresponds to energies several order of magnitude smaller than the typical
energy of the interactions. [After \protect\cite{valla_arpes_fermi_liquid_Mo}]}
\end{figure}

In addition, a remarkable experimental technique to look at the
single particle excitations, and momenta distributions is
provided by the photoemission technique
\cite{damascelli_review_ARPES}. Pending some hypothesis this
technique is a direct measure of the spectral function
$A(\vk,\omega)$ which is the probability of finding an
excitation with the energy $\omega$ and a momentum $\vk$. For
free particles $A(\vk,\omega) = \delta(\omega-\xi(\vk))$.
Naively one would expect that, because an energy of the order
of the interaction can be exchanged, these perfect peaks are
broadened over an energy of the order of the interaction. As
can be seen from \fref{fig:photo_exp}
this is clearly not the case. Very sharp peaks exist, and
become sharper and sharper as one gets closer to the Fermi
energy. The momentum distribution seems to be broadened
uniquely by the temperature when one is at the Fermi surface.

One is thus faced with a remarkable puzzle: the ``free
electron'' picture seems, at least qualitatively, to work
\emph{much better} than it should, based on estimates of the
interaction strength. This must hide a profound effect, and is
thus a great theoretical challenge.

\subsection{Landau Fermi liquid theory}

The solution to this puzzle was given by Landau and is known
under the name of Fermi liquid theory
\cite{landau_fermiliquid_theory_static,landau_fermiliquid_theory_dynamics}.

\begin{figure}
\begin{center}
\includegraphics[width=0.3\linewidth]{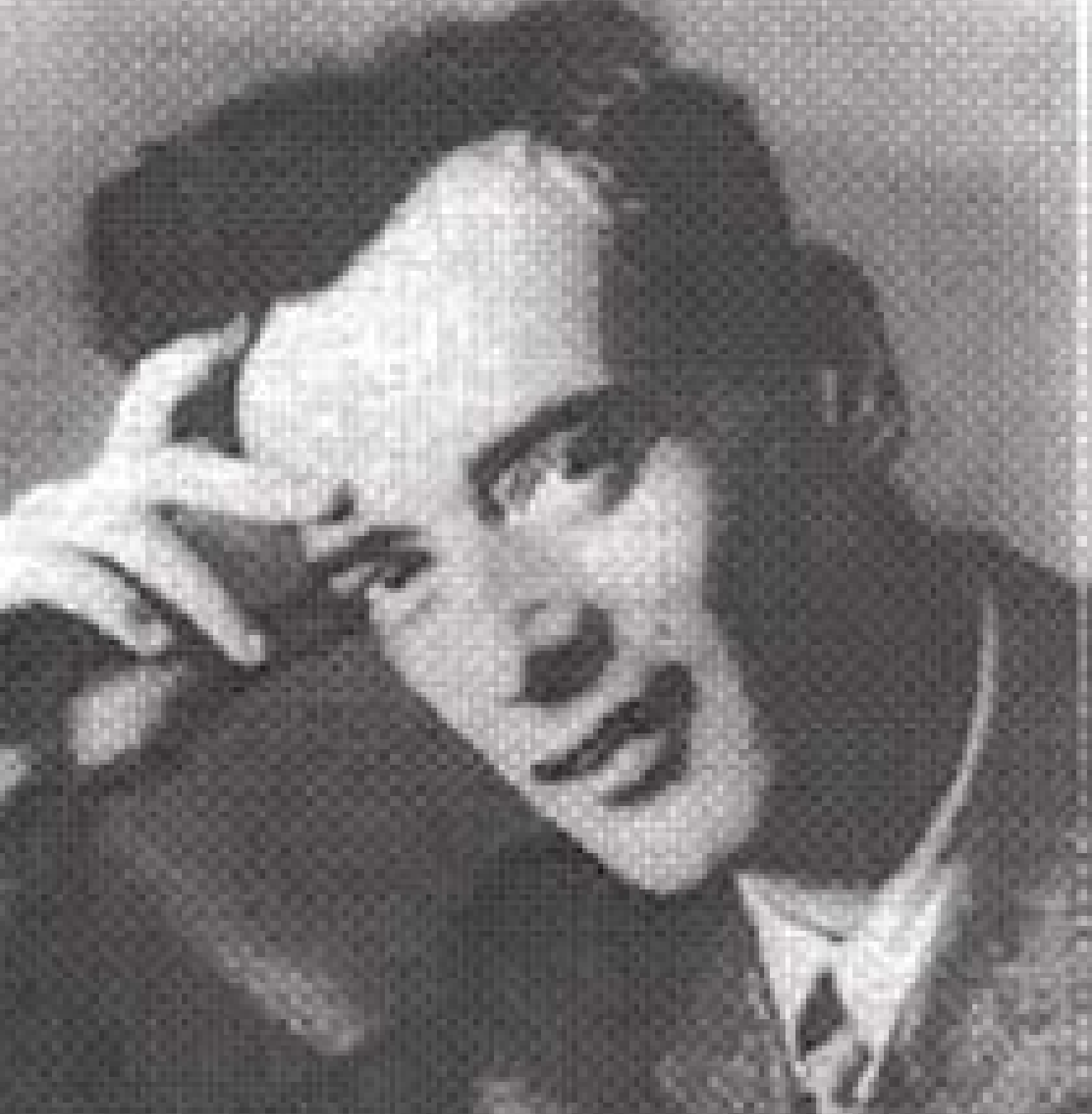}
\includegraphics[width=0.7\linewidth]{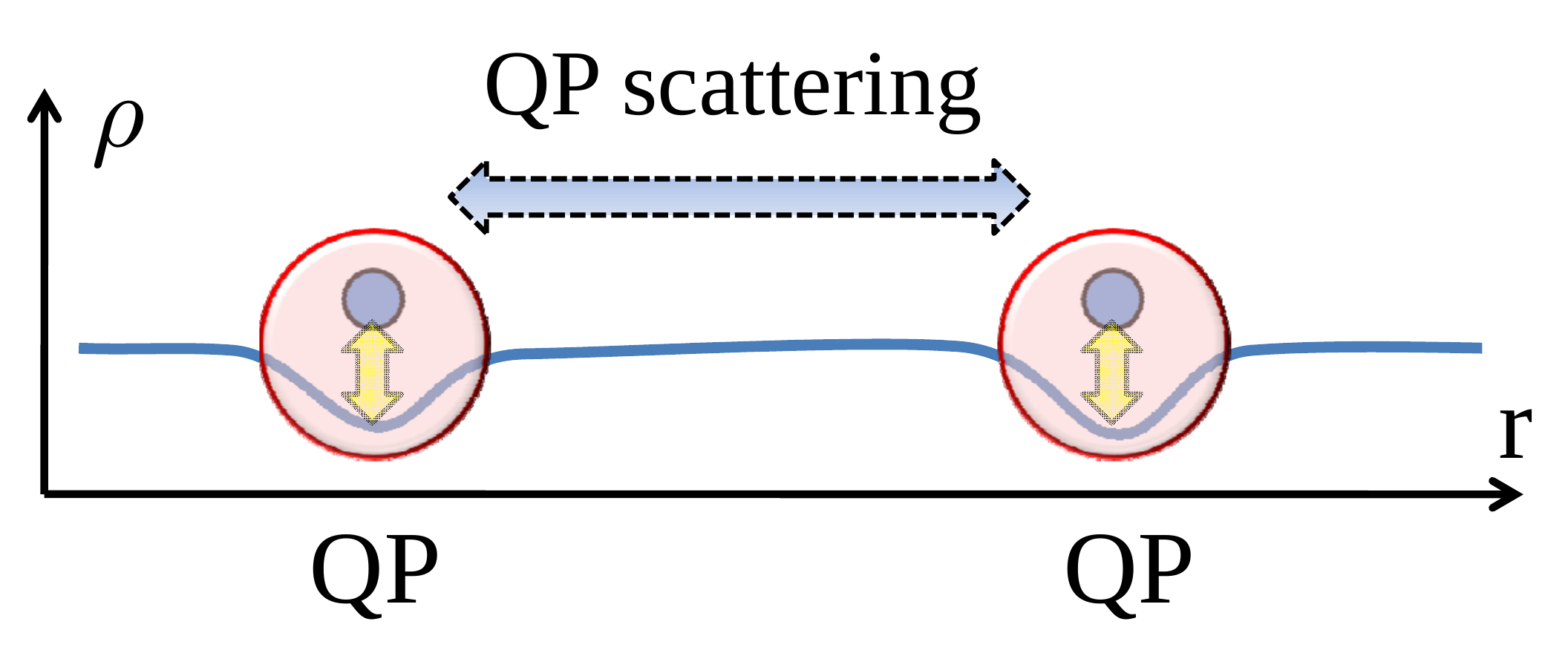}
\end{center}
\caption{\label{fig:fermiliquid} (top) Lev Landau, the man behind the Fermi liquid theory (and many other things).
(bot) In a Fermi liquid, the few excitations above the ground
state can interact strongly with all the other electrons
present in the ground state. The effect of such interactions is
strong and lead to a strong change of the parameters compared
to free electrons. The combined object behaves as a single
particle, named a quasiparticle. The quasiparticles have thus
characteristics depending strongly on the interactions. However
the scattering of the quasiparticles is blocked by the Pauli
principle leaving a very small phase space of scattering. The
lifetime of the quasiparticles is thus extremely large. This is
the essence of the Fermi liquid theory.}
\end{figure}

Let me first give a very qualitative description of the
underlying ideas before embarking on a more rigorous definition
and derivation. The main idea behind Fermi liquids
\cite{Nozieres_book} is to look at the excitations that would
exist above the ground state of the system. In the absence of
interactions the ground state is the Fermi sea
(\ref{eq:fermi_sea}). Let us assume now that one turns on the
interactions. This ground state will evolve into a very
complicated object, that we will be unable to describe, but
that does not interest us directly. What we need are the
excitations that correspond to the addition or removal
(creation of a hole) of a fermion in the ground state. In the
absence of interactions one just add an electron in an empty
$\vk$ state and such an excitation does not care about the
presence of all the other electrons in the ground state
(otherwise than via the Pauli principle which prevents from
creating it in an already occupied state). In the presence of
interactions this will not be the case and the added particle
interacts with the existing particles in the ground state. For
example for repulsive interactions one can expect that this
excitation repels other electrons in its vicinity. This is
schematically represented in \fref{fig:fermiliquid}.
On the other hand if one is at low temperature (compared to the
Fermi energy) there are very few of such excitations and one
can thus neglect the interactions between them. This picture
strongly suggests that the main interaction is between the
excitation and the ground state. This defines a new composite
object (fermion or hole surrounded by its own polarization
cloud). This complex object essentially behaves as a particle,
with the same quantum numbers (charge, spin) than the original
fermion, albeit with renormalized parameters, for example its
mass. This image thus strongly suggests that even in the
presence of interactions good excitations looking like free
particles, still exists. These particle resemble free fermions
but with a renormalized energy $E(\vk)$ and thus a renormalized
mass. Since the interaction has been incorporated in the
definition of such objects, it will not act as a source of
broadening for their momentum distribution and the momentum
distribution for the quasiparticles will remain very sharp,
with only the small temperature broadening.

\begin{figure}
 \begin{center}
 \includegraphics[width=0.6\linewidth]{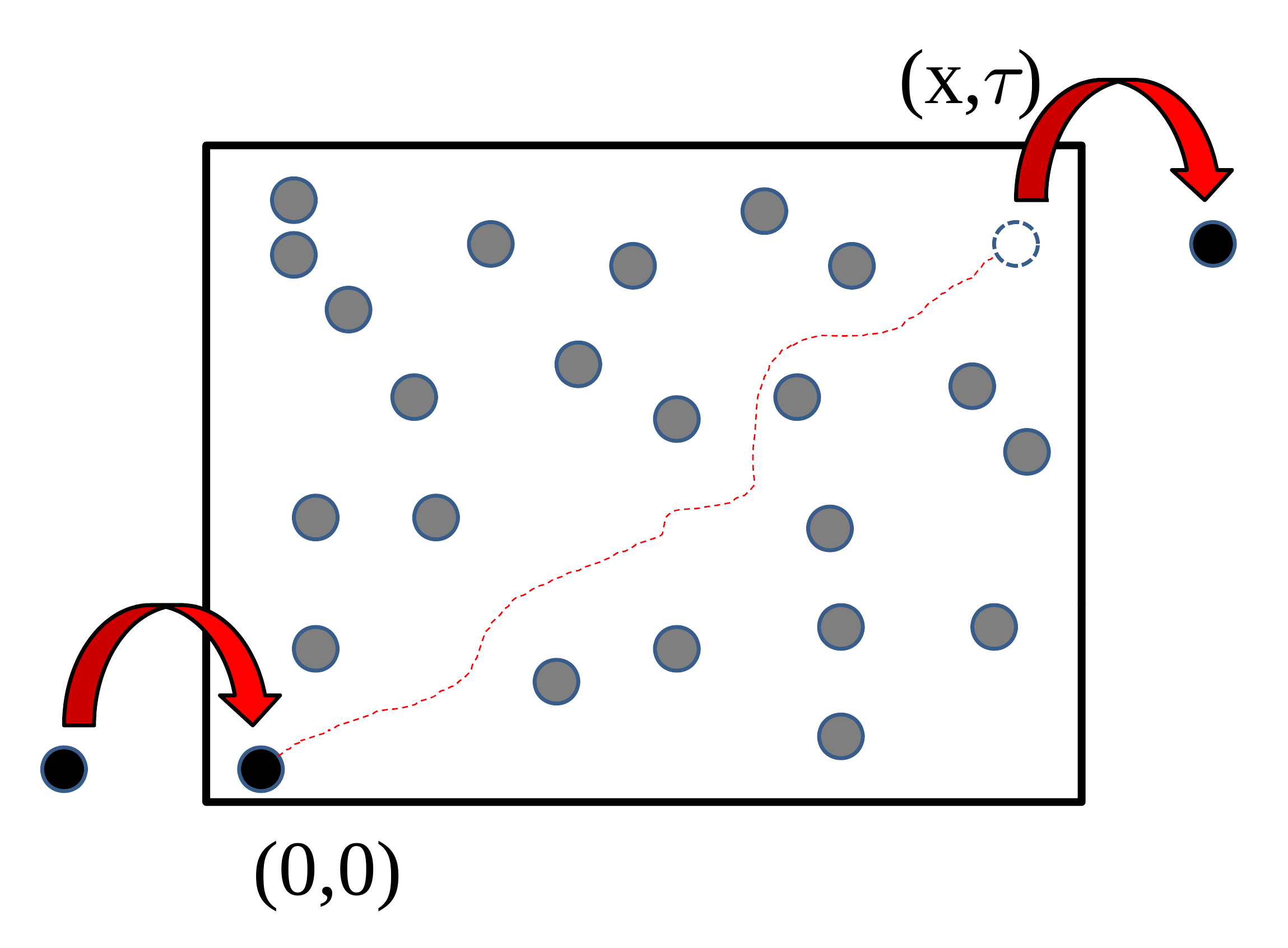}
 \end{center}
 \caption{\label{fig:gedankensingle}
 One way to understand the nature of the excitations in an interacting system is to inject a particle at point $r_1$ and time $t=0$.
 The resulting system evolves, and then one destroys a particle at time $t$ and point $r_2$. The amplitude of such a process indicates
 how much the propagation of the excitation between the two points and times is similar to the case of non interacting electrons or not.}
\end{figure}

Of course the above is just a qualitative idea. Let me now give
a more formal treatment. For that we can consider the retarded
single correlation function \cite{mahan_book,abrikosov_book}
\begin{equation} \label{eq:greendef}
 G(\vk,t_2-t_1) = - i \step(t_2-t_1) \langle \acomm{c_{\vk,t_2}}{\hc{c}_{\vk,t_1}} \rangle
\end{equation}
This correlation represents the creation of a particle in a
well defined momentum state $\vk$ at time $t_1$, let it
propagate and then tries to destroy it in a well defined
momentum state $\vk$ at time $t_2$. It thus measures how well
in the interacting system the single particle excitations still
resemble plane or Bloch waves, i.e. independent particles. This
corresponds to the gedanken experiment described in
\fref{fig:gedankensingle}.
The imaginary part of the Fourier transform of this correlation
function is just the spectral function
\begin{equation}
 A(\vk,\omega) = \frac{-1}{\pi} \Im G(\vk,\omega)
\end{equation}
The spectral function measures the probability to find a single
particle excitation with an energy $\omega$ and a momentum
$\vk$. It does obey the sum rule of probabilities $\int d\omega
A(\vk,\omega) = 1$. The general form of the retarded
correlation is
\begin{equation} \label{eq:fullsingle}
 G(\vk,\omega) = \frac1{\omega - \xi(\vk) - \Sigma(\vk,\omega) + i \delta}
\end{equation}
where $\Sigma(\vk,\omega)$, called the self energy, is a
certain function of momenta and frequency. The relation
(\ref{eq:fullsingle}) defines the function $\Sigma$. For
noninteracting systems $\Sigma = 0$, and perturbative methods
(Feynman diagrams) exist to compute $\Sigma$ in powers of the
interaction \cite{mahan_book,abrikosov_book}. However we will
not attempt here to compute the self energy $\Sigma$ but simply
to examine how it controls the spectral function. I incorporate
the small imaginary part $i \delta$ in $\Sigma$ for simplicity,
since one can expect in general $\Sigma$ to have a finite
imaginary part as well. The spectral function is
\begin{equation} \label{eq:fullspectral}
 A(\vk,\omega) = - \frac1\pi \frac{\Im\Sigma(\vk,\omega)}{(\omega-\xi(\vk)-\Re\Sigma(\vk,\omega))^2 + (\Im\Sigma(\vk,\omega))^2}
\end{equation}
We see that $\Im\Sigma$ and $\Re\Sigma$ play very different
roles in the spectral function. Note that quite generally
(\ref{eq:fullspectral}) imposes that $\Im\Sigma(\vk,\omega) <
0$ to get a positive spectral function. In the absence of
interactions $\Sigma = 0$ and one recovers
\begin{equation}
 A(\vk,\omega) = \delta(\omega - \xi(\vk))
\end{equation}

The imaginary part of the self energy gives the broadening of
the peaks, which now acquire a lorentzian like form, with a
width of order $\Im\Sigma$ and a height or order $1/\Im\Sigma$.
This can be readily seen from (\ref{eq:fullspectral}), for
example by setting the real part of $\Sigma$ to zero and taking
a constant $\Im\Sigma$. Note that now the peaks are \emph{not}
centered around $\xi(\vk)$ anymore but have a maximum at a an
energy $E(\vk)$ which is a solution $\omega = E(\vk)$ of
\begin{equation}
 \omega - \xi(\vk)-\Re\Sigma(\vk,\omega) = 0
\end{equation}
The imaginary part thus defines the broadening of the peaks,
i.e. how sharply the excitations are defined, while the real
part gives the \emph{new energy} of the excitations. Let us
look at these two elements in slightly more details.

\subsubsection{Lifetime:}

The imaginary part of the self energy controls the spread in
energy of the particles. This leads to the physical image of a
particle with an average energy $\omega = E(\vk)$, related to
its momentum, but with a certain spread $\Im\Sigma$ in energy.
To understand the physics of this spread let us consider the
Green function of a particle in real time:
\begin{equation} \label{eq:lifetime}
 G(\vk,t) = -i \step(t) e^{-i E(\vk) t} e^{-t/\tau}
\end{equation}
The oscillatory part is the normal time evolution of a
wavefunction of a particle with an energy $E(\vk)$. We have
added the exponential decay that would be produced by a finite
lifetime $\tau$ for the particle. The Fourier transform becomes
\begin{equation}
 G(\vk,\omega) = \frac{1}{\omega-E(\vk) + i/\tau}
\end{equation}
and the spectral function is
\begin{equation}
 A(\vk,\omega) = \frac1\pi\frac{1/\tau}{(\omega-E(\vk))^2 + (1/\tau)^2}
\end{equation}
which is essentially the one we are considering with the
identification
\begin{equation}
 \frac{1}{\tau} = \Im\Sigma
\end{equation}
We thus see from (\ref{eq:lifetime}) that a Lorentzian-like
spectral function corresponds to a particle with a well defined
energy $E(\vk)$ which defines the center of the peak, but also
with a finite \emph{lifetime} $\tau$. Of course the existence
of such lifetime does not mean that the particle physically
disappears, but simply that it does not exist as an excitation
with the given quantum number $\vk$. This is indeed an expected
effect of the interaction since the particle exchanges momentum
with the others particles and thus is able to change its
quantum state.

With the more general form of the self energy, which depends on
$\vk$ and $\omega$ this interpretation in terms of a lifetime,
still holds if the peak is narrow enough. Indeed in that case
the self energy at the position of the peak
$\Im\Sigma(\vk,\omega = E(\vk))$ matters if one assumes that
the self energy varies slowly enough with $\omega$ compared to
$\omega-E(\vk)$.

\subsubsection{Effective mass and quasiparticle weight:}

Let us now turn to the real part. For simplicity, let me set
the imaginary part to zero in (\ref{eq:fullspectral}) since in
this section we will be mostly interested by the position and
weight of the peak. This simplification replaces the Lorentzian
peaks by sharp $\delta$ functions, but keep the other
characteristics unchanged. With this simplification the
spectral function becomes
\begin{equation} \label{eq:specreal}
 A(\vk,\omega) = \delta(\omega - \xi(\vk) -\Re\Sigma(\vk,\omega))
\end{equation}
As we already pointed out, the role of the real part of the
self energy is to modify the position of the peak. One has now
a new dispersion relation $E(\vk)$ which is defined by
\begin{equation} \label{eq:disper}
 E(\vk) - \xi(\vk) - \Re\Sigma(\vk,\omega=E(\vk)) = 0
\end{equation}
The interactions, via the real part of the self-energy are thus
leading to a modification of the energy of single particle
excitations. Although we can in principle compute the whole
dispersion relation $E(\vk)$, in practice we do not need it
since the low energy excitations close to the Fermi level
control all the physical properties. At the Fermi level the
energy, with a suitable subtraction of the chemical potential,
is zero. One can thus expand it in powers of $\vk$. For free
electrons with $\xi(\vk) = \frac{\vk^2}{2m} - \frac{\kF^2}{2m}$
the corresponding expansion would give
\begin{equation} \label{eq:freeexp}
 \xi(\vk) = \frac{\kF}{m}(\vk-k_F)
\end{equation}
A similar expansion for the new dispersion $E(\vk)$ gives
\begin{equation}
 E(\vk) = 0 + \frac{\kF}{m^*} (\vk -\kF)
\end{equation}
which defines the coefficient $m^*$. Comparing with
(\ref{eq:freeexp}) we see that $m^*$ has the meaning of a mass.
Close to the Fermi level we only need to compute the effective
mass $m^*$ to fully determine (at least for a spherical Fermi
surface) the effects of the interactions on the energy of
single particle excitations. To relate the effective mass to
the self energy one computes from (\ref{eq:disper})
\begin{equation}
 \frac{d E(k)}{d k} = \frac{d \xi(k)}{d k} + \left.\frac{\partial \Re\Sigma(k,\omega)}{\partial k}\right|_{\omega=E(k)} +
                                      \left.\frac{\partial \Re\Sigma(k,\omega)}{\partial \omega}\right|_{\omega=E(k)} \frac{d E(k)}{d k}
\end{equation}
which can be solved to give
\begin{equation}
 \frac{\kF}{m^*} = \frac{\frac{\kF}{m} + \left.\frac{\partial \Re\Sigma(k,\omega)}{\partial k}\right|_{\omega=E(k)}}
               {1-\left.\frac{\partial \Re\Sigma(k,\omega)}{\partial \omega}\right|_{\omega=E(k)}}
\end{equation}
or in a more compact form
\begin{equation} \label{eq:effectivemass}
 \frac{m}{m^*} = \frac{1 + \frac{m}{\kF}\left.\frac{\partial \Re\Sigma(k,\omega)}{\partial k}\right|_{\omega=E(k)}}
               {1-\left.\frac{\partial \Re\Sigma(k,\omega)}{\partial \omega}\right|_{\omega=E(k)}}
\end{equation}
To determine the effective mass these relations should be
computed on the Fermi surface $E(\kF)=0$. The equation
(\ref{eq:effectivemass}) indicates how the self energy changes
the effective mass of the particles. This renormalization of
the mass by interaction is well consistent with the
experimental findings showing that in the specific heat one had
something that was resembling the behavior of free electrons
but with a different mass $m^*$.

However the interactions have another more subtle effect.
Indeed if we try to write the relation (\ref{eq:specreal}) in
the canonical form $\delta(\omega-E(\vk))$ that we would
naively expect for a free particle with the dispersion $E(\vk)$
one obtains from (\ref{eq:specreal})
\begin{equation}
 A(k,\omega) = Z_k \delta(\omega-E(k))
\end{equation}
with
\begin{equation} \label{eq:qpweight}
\begin{split}
 Z_k &= \left[\left.\frac{\partial }{\partial \omega}(\omega - \xi(k) - \Re\Sigma(k,\omega))\right|_{\omega=E(k)}\right]^{-1} \\
     &= \frac{1}{1 - \left.\frac{\partial \Re\Sigma(k,\omega)}{\partial\omega}\right|_{\omega=E(k)}}
\end{split}
\end{equation}
Because of the \emph{frequency} dependence of the real part of
the self energy, the total spectral weight in the peak is not
one any more one, but the total weight is now $Z_k$, which is
in general a number smaller than one. It is as if not the whole
electron (or rather the total spectral weight of an electron)
is converted into something that looks like a free particle
with a new dispersion relation, but only a faction $Z_k$ of it.
With our crude approximation the rest of the spectral function
has totally vanished. The fact that this violates the
conservation of the probability to find an excitation is
clearly an artefact of setting only the imaginary part to zero,
while keeping the real part, since real and imaginary part of
the self energy are related by a Kramers-Kronig relation.
However the reduction of the quasiparticle weight that we found
is quite real. What becomes of the remaining spectral weight
will be described in the next section.

To conclude we see that the real part of the self energy
controls the dispersion relation and the total weight of
excitations which in the spectral function produce peaks
exactly like free particles. The frequency and momentum
dependence of the real part of the self energy lead to the two
independent quantities $m^*$ the effective mass of the
excitations and $Z_k$ the weight. In the particular case when
the momentum dependence of the self energy is small on can see
from (\ref{eq:qpweight}) and (\ref{eq:effectivemass})
\begin{equation}
 \frac{m}{m^*} = Z_{\kF}
\end{equation}

\subsubsection{Landau Quasiparticles:}

\begin{figure}
\begin{center}
 \includegraphics[width=\columnwidth]{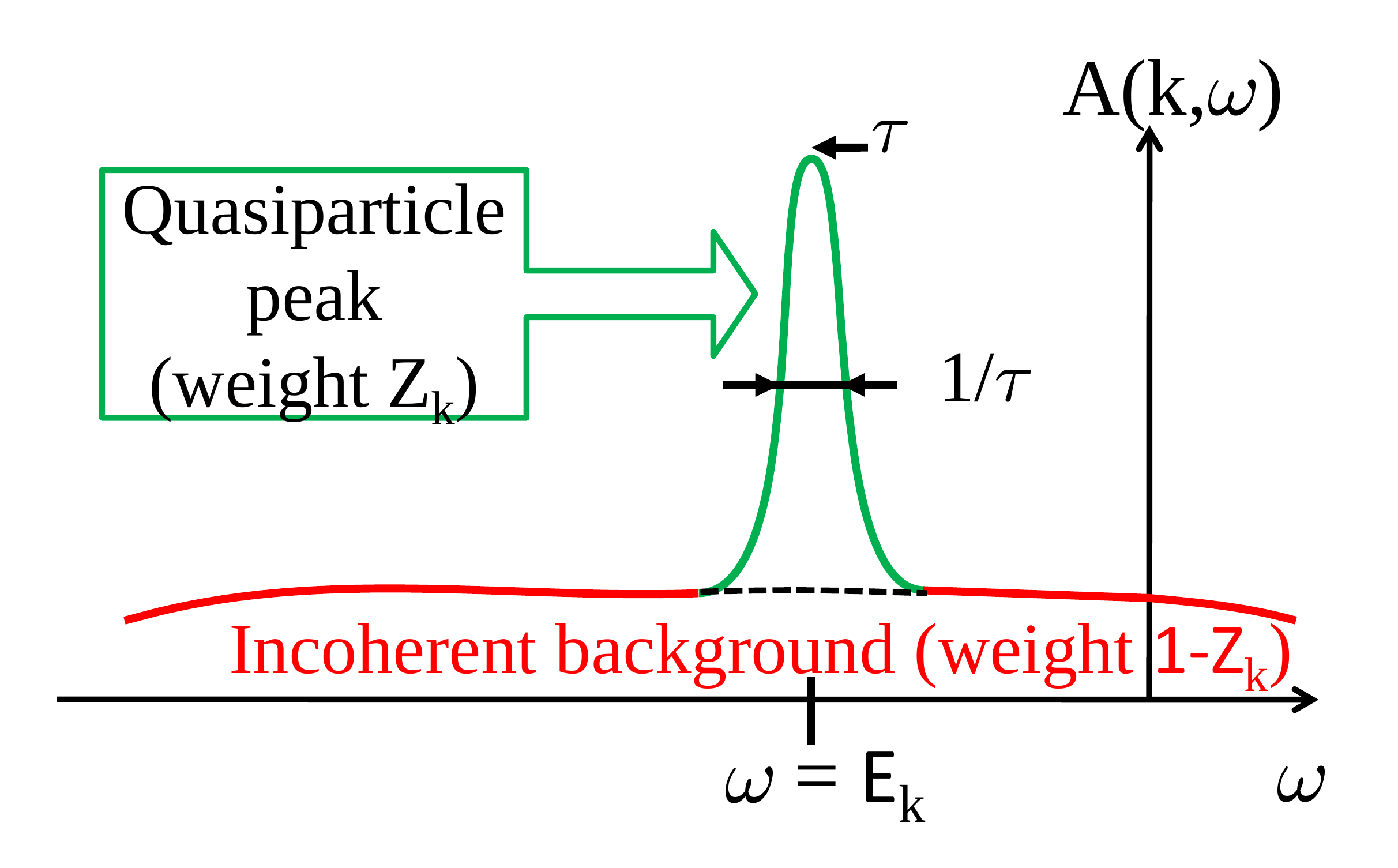}
\end{center}
\caption{\label{fig:spectralqp} A cartoon of the spectral function for interacting particle. One can recognize several features. There is a continuous background of excitations of total weight
$1-Z_k$. This part of the spectrum cannot be identified with excitations that resemble quasi-free particles. In addition to this continuous background there can be a quasiparticle peak. The
total weight of the peak is $Z_k$ determined by the real part of the self energy. The center of the peak is at an energy $E(k)$ which is renormalized by the interactions compared to the independent
electron dispersion $\xi(k)$. This change of dispersion defines an effective mass $m^*$ determined also by the real part of the self energy. The quasiparticle peak has a lorentzian lineshape that
traduces the finite lifetime of the quasiparticles. The lifetime is inversely proportional to the imaginary part of the self energy.}
\end{figure}

From the previous analysis of the spectral function and its
connection with the self energy we have a schematic idea of the
excitations as summarized in \fref{fig:spectralqp}.
Quite generally we can thus distinguish two parts in the
spectral function. There is a continuous background, without
any specific feature for which the probability to find a
particle with an energy $\omega$ is practically independent of
its momentum $k$. This part of the spectrum cannot be easily
identified with excitations resembling free or quasi-free
particles. On the other hand, in addition to this part, which
carries a total spectral weight $1-Z_k$, another part of the
excitations gives a spectral weight with a lorentzian peak,
well centered around a certain energy $E(k)$. This part of the
spectrum can thus be identified with a ``particle'', called
Landau quasiparticle, with a well defined relation between its
momentum $k$ and energy $\omega=E(k)$. This quasiparticle has a
only a finite lifetime, determined by the inverse width and
height of the peak. The dispersion relation and the total
weight of the quasiparticle peak are controlled by the real
part of the self energy, while the lifetime is inversely
proportional to the imaginary part. Depending on the self
energy, and thus the interactions, we can still have objects
that we could identify with ``free'' particles, solving our
problem of why the free electron picture works qualitatively so
well with just a renormalization of the parameters such as the
mass into an effective mass.

\begin{figure}
\begin{center}
 \includegraphics[width=\columnwidth]{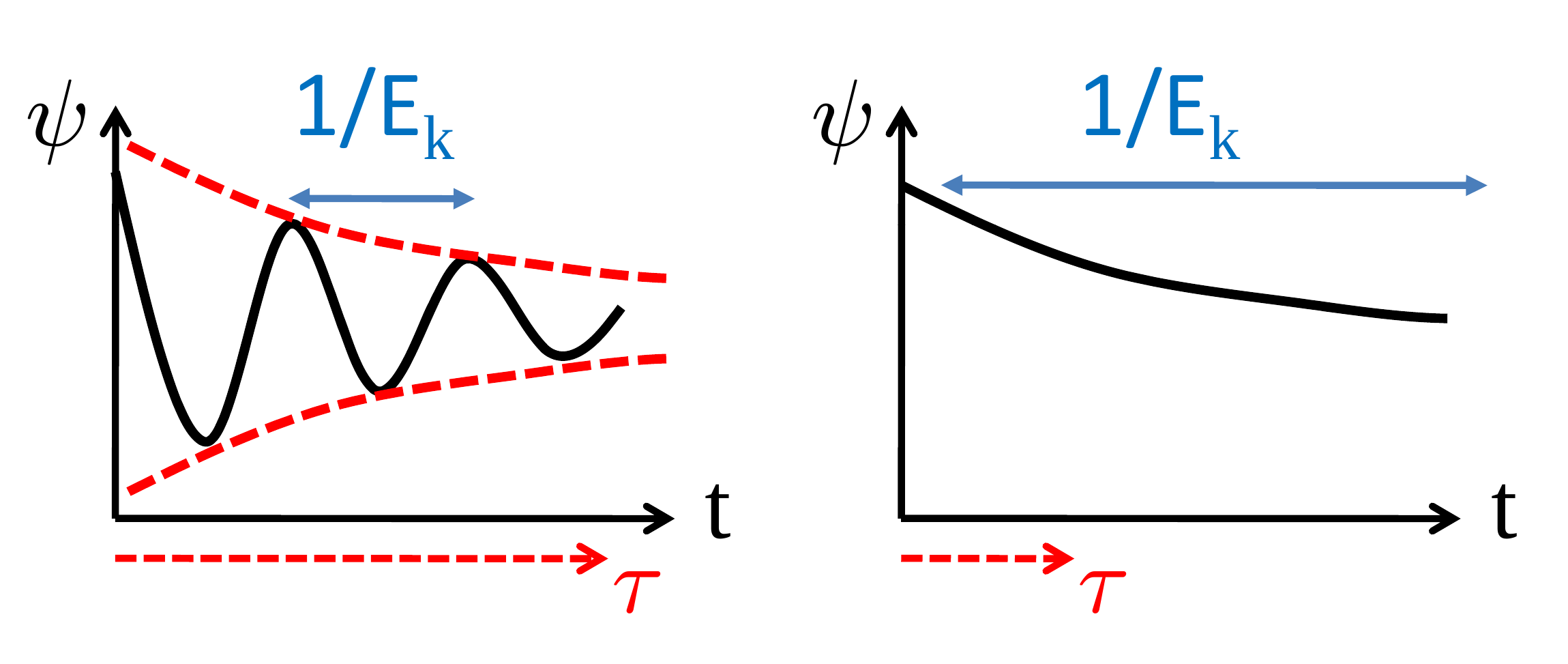}
\end{center}
\caption{\label{fig:amortiss} For particles with an energy $E(k)$ and a finite lifetime $\tau$, the energy controls the oscillations in time of the wavefunction.
(left) In order to properly identify an excitation as a particle it is mandatory that the wavefunction can oscillate several time before being damped by the lifetime, otherwise it is impossible
to precisely define the frequency of the oscillations. This is illustrated on the left part of the figure. (right) On the contrary if the damping is too fast, one cannot define an average energy
and thus identify the excitation with a particle.}
\end{figure}

However it is not clear that in the presence of interactions
one can have sharp quasiparticles. In fact one would naively
expect exactly the opposite. Indeed we would like to identify
the peak in the spectral function with the existence of a
quasiparticle. The energy of this excitation is $E(k)$ which of
course tends towards zero at the Fermi level, while the
imaginary part of the self energy is the inverse lifetime
$1/\tau$. Since $E(k)$ gives the oscillations in time of the
wavefunction of the particle $e^{- i E(k) t}$, in order to be
able to identify properly a particle it is mandatory, as shown
in \fref{fig:amortiss}
that there are many oscillations by the time the lifetime has
damped the wavefunction. This imposes
\begin{equation} \label{eq:qpcond}
 E(k)^{-1} \gg \tau
\end{equation}
Since $1/\tau$ is the imaginary part of the self energy and
controlled by energy scales of the order of the interactions,
one would naively expect the life time to be roughly constant
close to the Fermi level. On the other hand one has always
$E(k) \to 0$ when $k\to \kF$, and thus the relation
(\ref{eq:qpcond}) to be violated when one gets close to the
Fermi level. This would mean that for weak interactions one has
perhaps excitations that resemble particles far from the Fermi
level, but that this becomes worse and worse as one looks at
low energy properties, with finally all the excitations close
to Fermi level being quite different from particles. Quite
remarkably, as was first shown by Landau, this ``intuitive''
picture is totally incorrect and the lifetime has a quite
different behavior when one approaches the Fermi level.

\subsubsection{Quasiparticle scattering:}

\begin{figure}
\begin{center}
 \includegraphics[width=\columnwidth]{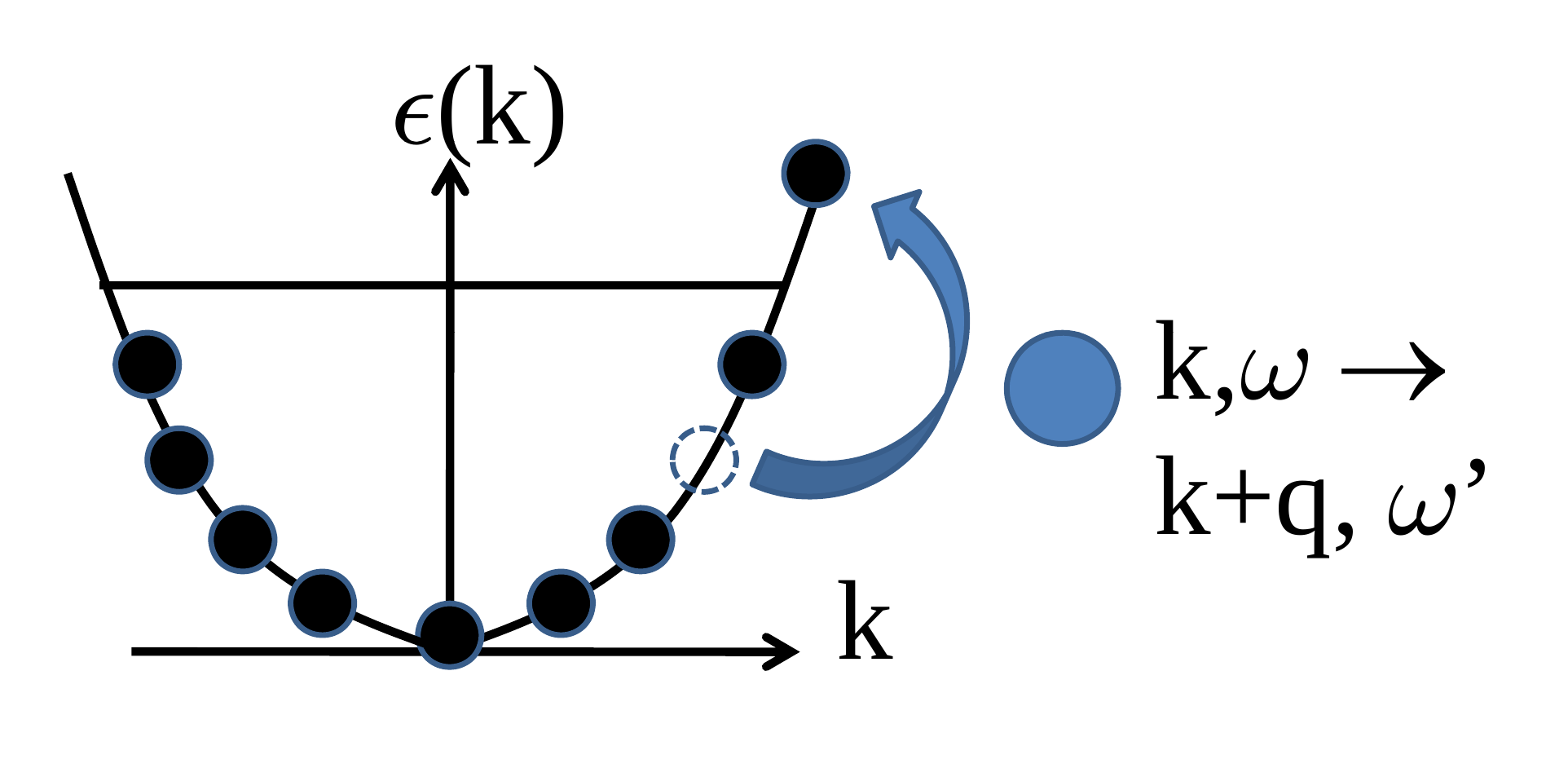}
\end{center}
\caption{\label{fig:parthole} Cartoon of the process giving the lifetime of a particle with energy $\omega$. It can interact with the ground state of the system, which has all single particle states filled below
the Fermi energy $\EF$. The excitations are thus particle hole excitations where a particle is promoted from below the Fermi level to above the Fermi level. Due to the presence of the sharp Fermi
level, the phase space available for making such particle hole excitations is severely restricted.}
\end{figure}

In order to estimate the lifetime let us look at what
excitations can lead to the scattering of a particle from a
state $k$ to another state. Let us start from the non
interacting ground state in the spirit of a perturbative
calculation in the interactions. As shown in
\fref{fig:parthole}
a particle coming in the system with an energy $\omega$ and a
momentum $k$ can excite a particle-hole excitation, taking a
particle below the Fermi surface with an energy $\omega_1$ and
putting it above the Fermi level with an energy $\omega_2$. The
process is possible if the initial state is occupied and the
final state is empty. One can estimate the probability of
transition using the Fermi golden rule. The probability of the
transition gives directly the inverse lifetime of the particle,
and thus the imaginary part of the self energy. We will not
care here about the matrix elements of the transition, assuming
that all possible transitions will effectively happen with some
matrix element. The probability of transition is thus the sum
over all possible initial states and final states that respect
the constraints (energy conservation and initial state
occupied, final state empty). Since the external particle has
an energy $\omega$ it can give at most $\omega$ in the
transition. Thus $\omega_2 - \omega_1 \leq \omega$. This
implies also directly that the initial state cannot go deeper
below the Fermi level than $\omega$ otherwise the final state
would also be below the Fermi level and the transition would be
forbidden. The probability of transition is thus
\begin{equation}
 P \propto \int_{-\omega}^{0} d\omega_1 \int_0^{\omega+\omega_1} d\omega_2 = \frac12 \omega^2
\end{equation}
One has thus the remarkable result that because of the
\emph{discontinuity} due to the Fermi surface and the Pauli
principle that only allows the transitions from below to above
the Fermi surface, the inverse lifetime behaves as $\omega^2$.
This has drastic consequences since it means that contrarily to
the naive expectations, when one considers a quasiparticle at
the energy $\omega$, the lifetime grows much \emph{faster} than
the period $D \sim 1/\omega$ characterizing the oscillations of
the wavefunction. In fact
\begin{equation}
\frac{\tau}{D} = \frac{1}{\omega} \to \infty
\end{equation}
when one approaches the Fermi level. In other words the Landau
quasiparticles become \emph{better and better defined} as one
gets closer to the Fermi level. This is a remarkable result
since it confirms that we can view the system as composed of
single particle excitations that resemble the original
electrons, but with renormalized parameters (effective mass
$m^*$ and quasiparticle weight $Z_k$). Other quantum numbers
are the same than the ones of an electron (charge, spin). Note
that this does \emph{not} mean that close to the Fermi level
the interactions are disappearing from the system. They are
present and can be extremely strong, and affect both the
effective mass and quasiparticle weight very strongly. It is
only the scattering of the quasiparticles that is going to zero
when one is going close to the Fermi level. This is thus a very
unusual situation quite different from what would happen in a
classical gas. In such a case diluting the gas would thus
reduce both the interaction between the particles and also
their scattering in essentially the same proportion. On the
contrary in a Fermi liquid there are many $\npart \to \infty$
electrons in the ground state, which are in principle strongly
affected by the interactions. Note again that computing the
ground state would be a very complicated task. However there
are very few excitations above this ground state at low energy.
These excitations can interact strongly with the other
electrons in the soup of the ground state, leading to a very
strong change of the characteristics compared to free electron
excitations. This can lead to very large effective masses or
small quasiparticle weight. On the other hand the lifetime of
the quasiparticles is controlled by a totally different
mechanism since it is blocked by the Pauli principle, as shown
in \fref{fig:fermiliquid}. Thus even if the interaction is
strong the \emph{phase space} available for such a scattering
is going to zero close to the Fermi level, making the
quasiparticle in practice infinitely long lived particles, and
allowing to use them to describe the system. The image of
\fref{fig:parthole} also gives us a description of what a
quasiparticle is: this is an electron that is surrounded by a
cloud of particle-hole excitations, or in other words density
fluctuations since $\hc{c}_{k+q}c_k$ is typically the type of
terms entering the density operator. Such density fluctuations
are of course neutral and do not change the spin. This
composite object electron+density fluctuation cloud, thus
represent a tightly bound object (just like an electron does
dress with a cloud of photons in quantum electrodynamics), that
is the Landau quasiparticle. Since the electron when moving
must carry with it its polarization cloud, one can guess that
its effective mass will indeed be affected.

\begin{figure}
\begin{center}
 \includegraphics[width=0.8\linewidth]{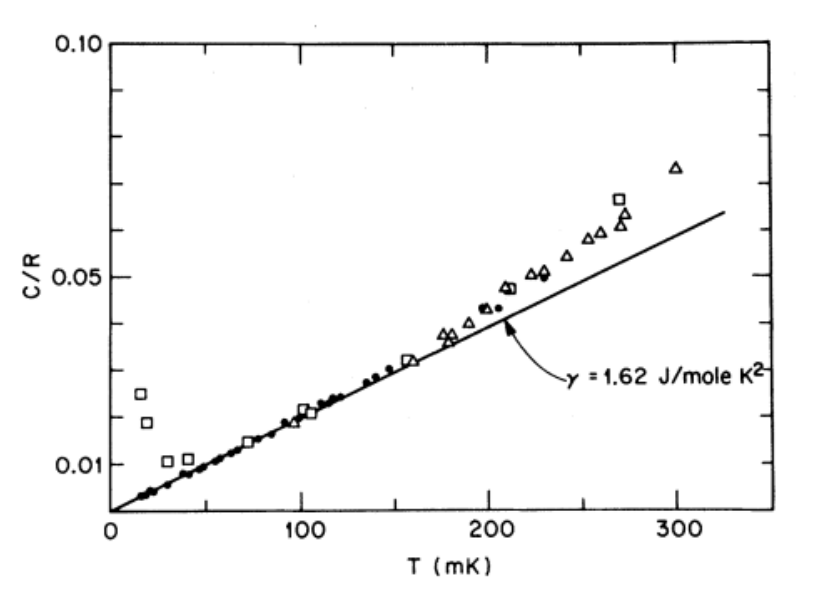}
 \includegraphics[width=0.6\linewidth]{Figures/heavy_fermions_1}
\end{center}
\caption{\label{fig:heavymass} Physical properties of the compound CeAl$_3$. (top) The specific
heat is linear in temperature $T$, but the slope gives an effective mass of about $10^3$m showing extremely
strong interaction effects, showing that the Fermi liquid theory applies even when the interaction effects are strong.
(bot) The resistivity varies as $T^2$ in very good agreement with the Fermi liquid theory.
[After \protect\cite{andres_CeAl3_effective_mass}]}
\end{figure}

The Fermi liquid theory is a direct explanation of the fact
that ``free'' electrons theory works very well
\emph{qualitatively} (such as the specific heat linear in
temperature) even when the change of parameters can be huge. We
show in \fref{fig:heavymass} the case of systems where the
renormalization of the mass is about $m^* \sim 10^3 m$
indicating very strong interactions effects.
Nevertheless we see that the specific heat varies linearly with
temperature just like for free electrons. The prediction for
the quasiparticle peaks fits very well with the photoemission
data of \fref{fig:photo_exp}, in which one clearly sees the
peaks becoming sharper as one approaches the Fermi level. There
is another direct consequence of the prediction for the
lifetime. At finite temperature one can expect the lifetime to
vary as $\tau \sim 1/T^2$ since $T$ is the relevant energy
scale when $T \gg \omega$. If we put such a lifetime in the
Drude formula for the conductivity we get
\begin{equation}
 \sigma(T) = \frac{n e^2 \tau}{m} \propto \frac{1}{T^2}
\end{equation}
This result can be confirmed by a full calculation. This shows
that the electron-electron interactions give an intrinsic
contribution to the resistivity that varies as $\rho(T) \sim
T^2$, and which also can be taken as one of the characteristic
of Fermi liquid behavior. This is however difficult to test
since this temperature dependence can easily be masked by other
scattering phenomena (impurities, scattering by the phonons
etc.) that must be added to the electron-electron scattering
and that have quite different temperature dependence.
Nevertheless there are some materials where the $T^2$ law can
be well observed as shown in \fref{fig:heavymass}. Another
interesting consequence can be deduced by looking at the
occupation factor $n(k)$ which can be expressed from the
spectral function by
\begin{equation}
 n(k) = \int d\omega' A(k,\omega') \Fermi(\omega')
\end{equation}
For free electrons one recovers the step function at $k=\kF$.
For a Fermi liquid, if we represent the spectral function as
\begin{equation}
 A(k,\omega) = Z_k \delta(\omega-E(k)) + A_{{\rm inc}}(k,\omega)
\end{equation}
where the incoherent part is a smooth flattish function without
any salient feature, then $n(k)$ becomes
\begin{equation}
 n(k) = Z_k \Fermi(E(k)) + Cste
\end{equation}
Thus even in the presence of interaction there is still a
\emph{discontinuity} at the Fermi level, that is only rounded
by the temperature. Contrarily to the case of free electron the
amplitude of the singularity at $T=0$ is not one anymore but is
now $Z_{\kF} < 1$. The existence of this discontinuity
\emph{if} quasiparticle exists tells us directly that the Fermi
liquid theory is internally consistent since the very existence
of the quasi particles (namely the large lifetime) was heavily
resting on the existence of such a discontinuity at the Fermi
level. One can thus in a way consider that the existence of a
sharp discontinuity at the Fermi level is a good order
parameters to characterize the existence of a Fermi liquid.

One important question is when the Fermi liquid theory does
apply. This is of course a very delicate issue. One can see
both from the arguments given above, and from direct
perturbative calculations that when the interactions are weak
the Fermi liquid theory will in general be valid. There are
some notable exceptions that we will examine in the next
section, and for which the phase space argument given above
fails. However the main interest of the Fermi liquid theory is
that it does not rest on the fact that the interactions are
small and, as we have seen through examples, works also
remarkably well for the case of strong interactions, even when
all perturbation theory fails to be controlled. This is
specially important for realistic systems since, as we showed,
the interaction is routinely of the same order than the kinetic
energy even in very good metals. The Fermi liquid theory has
thus been the cornerstone of our description of most condensed
matter systems in the last 50 years or so. Indeed it tells us
that we can ``forget'' (or easily treat) the main perturbation,
namely the interaction among fermions, by simply writing what
is essentially a free fermion Hamiltonian with some parameters
changed. It is not even important to compute microscopically
these parameters since one can simply extract them from one
experiment and then use them consistently in the others. This
allows to go much further and treat effects caused by much
smaller perturbations that one would otherwise have been
totally unable to take into account. One of the most
spectacular examples is the possibility to now look at the very
tiny (compared to the electron-electron interactions)
electron-phonon coupling, and to obtain from that the solution
to the phenomenon of superconductivity, or other instabilities
such as magnetic ordering.

\section{Beyond Fermi liquid}

Of course not all materials follow the Fermi liquid theory.
There are cases when this theory fails to apply. In that case
the system is commonly referred to as strongly correlated or
``non Fermi liquid'' a term that hides our poor knowledge of
their properties. For such systems, the question of the effects
of interactions becomes again a formidable problem. As
discussed in the introduction, most of the actual research in
now devoted to such non fermi liquid systems.

There are fortunately some situation where one can understand
the physics and we will examine such cases in this section as
well as define the main models that are at the heart of the
study of these systems.

\subsection{Instabilities of the FL}

The Fermi liquid can become unstable for a variety of reasons.
Some of them are well known.

The simplest instability consists, at low temperature, for the
system to go an ordered state. Many type of orders are
possible, the most common are spin order such as
ferromagnetism, antiferromagnetism, charge order such as a
charge density wave, or superconductivity. In general analyzing
such instabilities can be done by computing the corresponding
susceptibility and looking for divergences as the temperature
is lowered. When the normal system is well described by a Fermi
liquid it is in general relatively easy to compute these
susceptibilities by a mean field decoupling of the interaction.
This is of course much more complex when one starts from a
normal phase which is a non Fermi liquid.

Another important ingredient in the stability of the Fermi
liquid phase is the dimensionality of the system. Intuitively
we can expect that the lower the dimension the more important
the effects of the interactions will be since the particles
have a harder time to avoid each other. The ultimate case in
that respect is the one dimensional situation where one
particle moving will push the particle in front and so on, as
anybody queuing in a line had already has chance to notice.
This effect of dimensionality is confirmed by a direct
perturbative calculation of the self energy
\cite{mahan_book,abrikosov_book}
\begin{equation}
\begin{split}
 \Sigma_{3D}(\omega) & \propto U^2 \omega^2 \\
 \Sigma_{2D}(\omega) & \propto U^2 \omega^2\log(\omega) \\
 \Sigma_{1D}(\omega) & \propto U^2 \omega\log(\omega)
\end{split}
\end{equation}
where $U$ is the strength of the interaction. In particular one
immediately see that for the one dimensional case, the self
energy becomes dominant compared to the mean energy $\omega$.
Using (\ref{eq:qpweight}) one sees that for one dimension the
quasiparticle weight is zero at the Fermi level. This means the
whole argument we used in the previous section to justify the
existence of sharper and sharper peaks an thus the existence of
Landau quasiparticles fails \emph{regardless} of the strength
of the interaction. In one dimension Fermi liquid always fails.
This leads to a remarkable physics that we examine in more
details in \sref{sec:onedim}.

With the exception of the one dimensional case, one can thus
expect the Fermi liquid to be perturbatively valid in two
dimensions and above. Of course if the interaction becomes
large whether a non Fermi liquid state can appear is one of the
major challenges of today's research.

\subsection{Tight binding and Hubbard Hamiltonian}

One ingredient of special important is the presence of a
lattice on which the fermions can move. This is the normal
situation in solids, with the presence of the ionic lattice,
and can be realized very well in cold atomic systems by
imposing an optical lattice. Even for free electrons the
lattice is an essential ingredient, since it leads to the
existence of bands of energy.

One very simple description of the effects of a lattice is
provided by the tight binding model \cite{ziman_solid_book}.
This model is of course an approximation of the real band
structure in solids, but contains the important ingredients.
Recently cold atomic systems in optical lattices have provided
excellent realization of such a model, and we recall here its
main features.

Let us consider of cold atoms of fermions or bosons
\cite{anglin_review_optical}. Such system can be described by
\begin{equation} \label{eq:boscont}
 H = \int dr \frac{\hbar^2(\nabla\psi)^\dagger(\nabla\psi)}{2m} +
 \frac12\int dr \;dr'\; V(r-r') \rho(r)\rho(r') - \int dr\;\mu(r) \rho(r)
\end{equation}
The first term is the kinetic energy, the second term is the
interaction $V$ between the particles and the last term is the
chemical potential. The three dimensional interaction is
characterized by a scattering length \cite{pitaevskii_becbook}
$a_s$
\begin{equation} \label{eq:interatom}
 V(x,y,z) = V_0\delta(x)\delta(y)\delta(z) = \frac{4\pi\hbar^2a_s}{m}\delta(x)\delta(y)\delta(z)
\end{equation}
The chemical potential takes into account the confining
potential $V_c(r)$ keeping the particles in the trap
\begin{equation} \label{eq:varchem}
 \mu(\vr) = \mu_0 - V_c(r) = \mu_0 - \frac12 \omega_0 r^2
\end{equation}
and is thus in general position dependent. In presence of
interactions the effect of the confining potential can usually
be taken by making a local density approximation. If one
neglects the kinetic energy (so called Thomas-Fermi
approximation; for other situations see
\cite{pitaevskii_becbook}) the density profile is obtained by
minimizing (\ref{eq:boscont}) and (\ref{eq:varchem}), leading
to
\begin{equation}
 V_0 \rho(r) + [V_c(r)-\mu_0] = 0
\end{equation}
the density profile is thus an inverted parabola, reflecting
the change of the chemical potential. In the following we will
ignore for simplicity the confining potential and consider the
system as homogeneous.

If one adds an optical lattice, produced by counter propagating
lasers, to the system it produces a periodic potential of the
form $V_L(x)$ coupled to the density
\cite{bloch_cold_atoms_optical_lattices_review}
\begin{equation} \label{eq:boslat}
 H_L = \int dx\; V_L(x) \rho(x)
\end{equation}
This term, which favors certain points in space for the
position of the bosons, mimics the presence of a lattice of
period $a$, the periodicity of the potential $V_L(x)$. We take
the potential as
\begin{equation}
 V_L(x) = V_L \sum_{\alpha=1}^d \sin^2(k_a r_\alpha) = \frac{V_L}2 \sum_{\alpha=1}^d [1-\cos(2 k_\alpha r_\alpha)]
\end{equation}
one has thus $a = \pi/k_\alpha$ as the lattice spacing.

\begin{figure}
\begin{center}
 \includegraphics[width=\columnwidth]{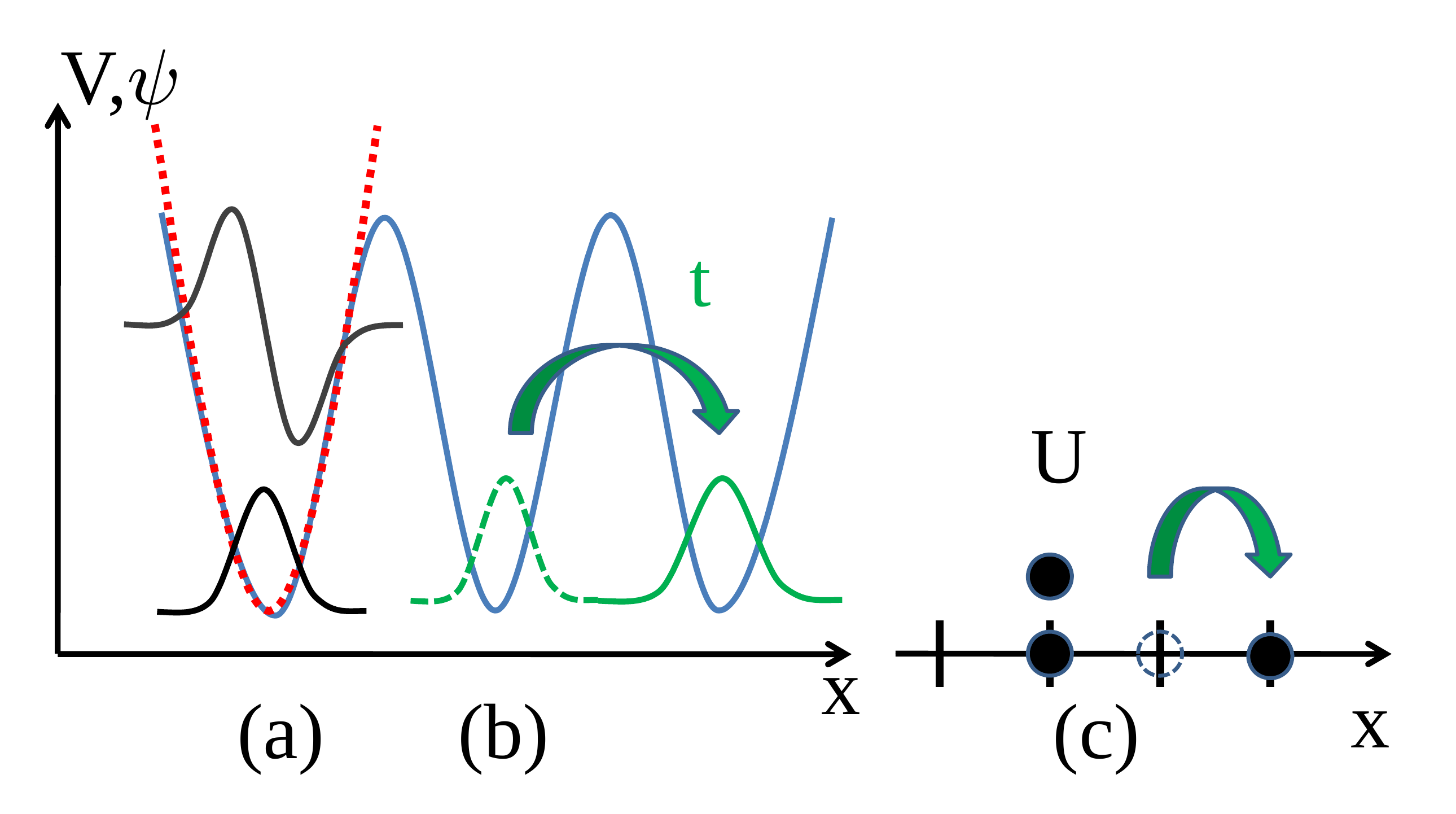}
\end{center}
\caption{\label{fig:optical} Optical lattices and tight binding model. (a) If the optical lattice is deep
enough one can consider the wells as practically independent and obtain the states in each well. Since the well can be approximated
by a parabolic potential one has harmonic oscillator states. (b) Some level of overlap exists between the wavefunctions of different
wells. This provide a hopping amplitude $t$ for a particle to move from one well to the next. Similarly the interaction between particles is only
efficient if the two particles are in the same well. (c) The system can be mapped on a tight binding model, describing particles leaving on the
sites of a lattice and hopping from one site to the next with an amplitude $t$. If interactions are included they can be modelled
by local interactions of strength $U$. The resulting Hamiltonian is the famous Hubbard model.}
\end{figure}

If the lattice amplitude $V_L$ is large then it is possible to
considerably simplify the full Hamiltonian. In that case on can
focuss on the solution in one of the wells of the optical
lattice, as shown in \fref{fig:optical}.

The well can be approximated by a parabola $\frac12 (4V_L) k^2
r^2$ and the solutions are those of an harmonic oscillator. The
ground state one is
\begin{equation} \label{eq:harmfond}
 \psi_0(x) = \left(\frac{m \omega_0}{\hbar \pi }\right)^{1/4} e^{-\frac{m \omega_0}{2 \hbar} x^2}
\end{equation}
and higher excited states are schematically indicated on
\fref{fig:optical}. Typical values for the above parameters are
$a_s \sim 5 nm$ while $a \sim 400 nm$
\cite{stoferle_tonks_optical}.

If $V_L$ is large the energy levels in each well are well
separated and one can retain only the ground state wavefunction
in each well, higher excited state being well separated in
energy. On can thus full represent the state of the system by a
creation (or destruction) of a particle in the $i$-th well
(i.e. with the wavefunction (\ref{eq:harmfond})). One has thus
the energy
\begin{equation}
 H = E_0 \sum_u \hc{c}_i c_i
\end{equation}
This is just a chemical potential term and can be absorbed in
the chemical potential. Of course if the lattice is not
infinite, as indicated on \fref{fig:optical} the wavefunction
in different wells overlap, and there is an hybridation term
$t$ between the wavefunction in two adjacent wells. This term
allows the particle to go from one well to the next. For an
optical lattice one gets \cite{zwerger_JU_expressions}
\begin{equation} \label{eq:opticalparam}
  t/E_r = (4/\sqrt{\pi}) (V_L/E_r)^{(3/4)} \exp{(-2\sqrt{V_L/E_r})}
\end{equation}
Here $E_r = \hbar^2 k^2/(2m)$ is the so called recoil energy,
i.e. the kinetic energy for a momentum of order $\pi/a$. In
presence of such term the energy becomes (we have written it
for two spin species)
\begin{equation}
 H_t = - t \sum_{\langle ij \rangle, \sigma} \hc{c}_{i\sigma} c_{j\sigma} - \sum_i \mu_{i\sigma} \hc{c}_{i\sigma} c_{i\sigma}
\end{equation}
where $\langle\rangle$ denotes nearest neighbors. In Fourier
space one recovers (\ref{eq:kinetic3}) with
\begin{equation} \label{eq:tight_disp}
 \xi(\vk) = - 2t \sum_{j=1}^d \cos(k_j a)
\end{equation}
where $d$ is the dimension of space, $a$ the lattice spacing (I
assumed here for simplicity a square lattice). The momentum
$\vk$ is restricted to the First Brillouin zone
\begin{equation}
 k_j \in [-\pi/a,\pi/a]
\end{equation}
Thus one particle per site corresponds to an half filled zone
(half of the available states are occupied). For example in one
dimension this gives $\kF = \pi/(2a)$. For independent fermions
this is the best situation to have a metallic state since the
Fermi level is far from a band edge. An empty band and a
completely filled band (two particles per site) correspond to
an insulating state. For a filled band all the excitations are
blocked by the Pauli principle. The tight binding description
of the kinetic energy contains thus the essential ingredient of
quantum periodic systems namely the existence of bands, and the
existence of a Brillouin zone.

One can add to this description, the effects of interaction. In
a solid, as we discussed, the interaction is screened. A good
approximation is thus to take a local interaction. For optical
lattice such an approximation is truly excellent. Indeed, since
atoms are neutral the interaction (\ref{eq:interatom}) has a
range much smaller than the size of the well. Thus atoms only
interact if they are in the same well of the optical lattice.
One can thus represent the interaction by a purely local term
\begin{equation}
 H_U = U \sum_i n_{i\up} n_{i\down}
\end{equation}
The effective potential $U$ can be easily computed  by using
the shape of the on site wave function (\ref{eq:harmfond}) with
$\omega_0^2 = 4 V_L k^2$
\begin{equation}
 U  = \int dxdydz |\psi(x,y,z)|^4
\end{equation}
where $\psi(x,y,z) = \psi_0(x)\psi_0(y)\psi_0(z)$.

The full Hamiltonian of the system is thus
\begin{equation} \label{eq:hubbard}
 H = H_t + H_U
\end{equation}
and describes particle hopping on a lattice with a purely local
repulsion. This is the famous Hubbard model. This model,
although extremely simple contains the essential ingredients of
interacting fermions. Optical lattices are thus excellent
realizations of this model. They offer in addition a powerful
control on the ratio of $U/t$ the interaction measured compared
to the kinetic energy, since we see that by increasing the
height of the optical lattice one essentially modifies the
overlap integral $t$ between two different sites
\cite{jaksch_bose_hubbard,greiner_mott_bec}. Recently cold
atomic systems have also provided a direct control over the
interaction $U$ by using a Feshbach resonance
\cite{bloch_cold_atoms_optical_lattices_review}.

\subsection{Mott insulators}

\begin{figure}
\begin{center}
\includegraphics[width=0.2\linewidth]{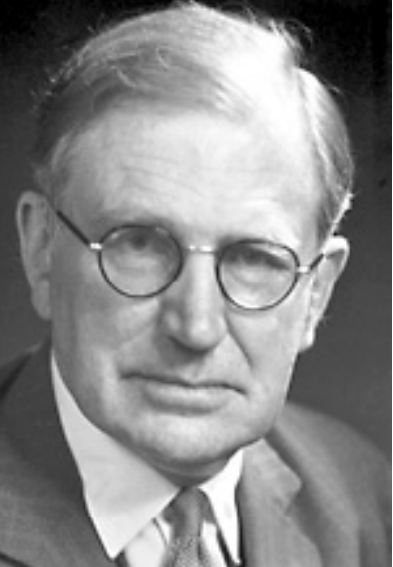}
\includegraphics[width=0.8\linewidth]{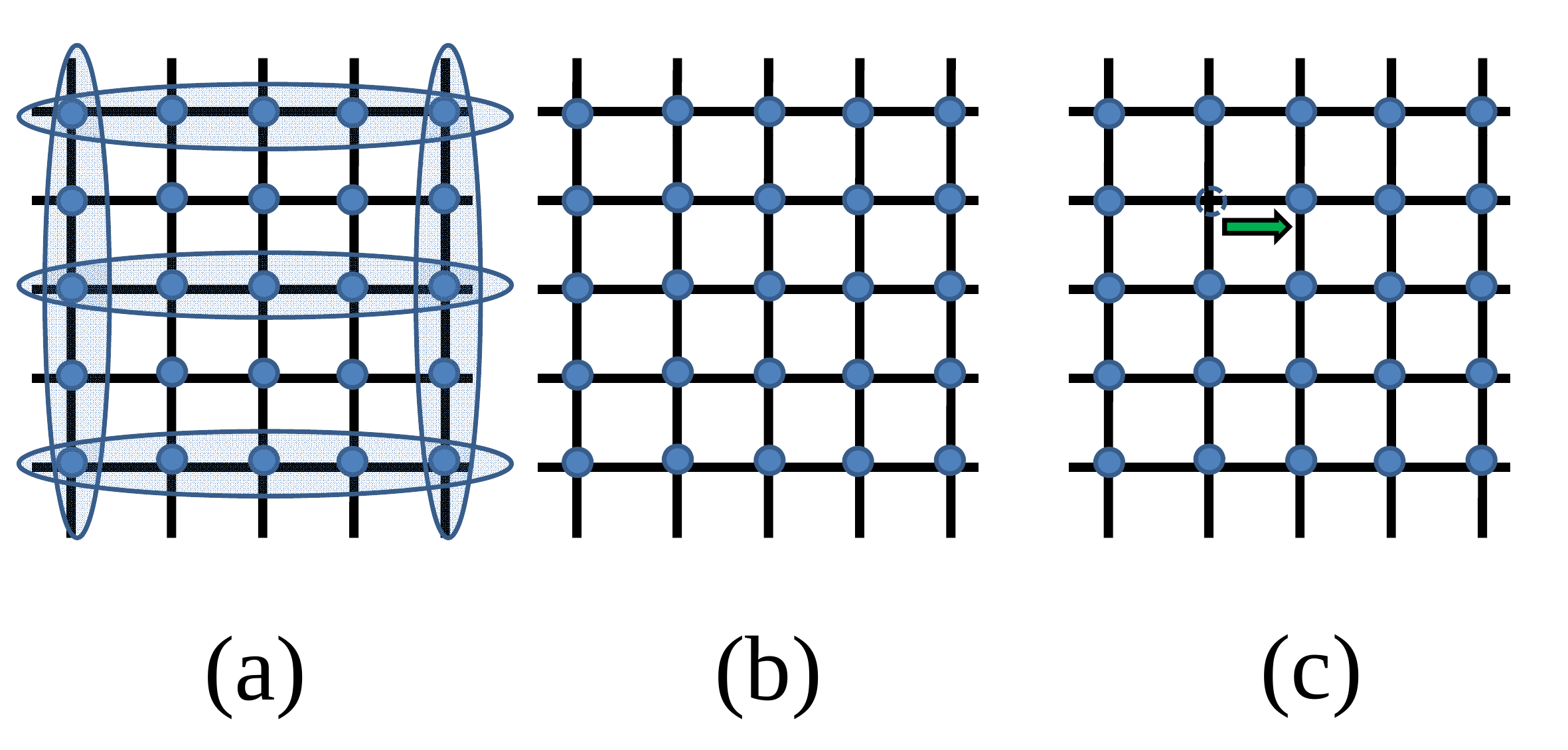}
\end{center}
\caption{\label{fig:mott} (top) Sir Nevil Mott. (a) If the repulsion is weak the particles prefer to gain
kinetic energy and are essentially in delocalized (plane wave) state. This leads to a metallic state, and usually to a
Fermi liquid. (b) For one particle per site $n=1$ and strong interactions the system prefers to localize the particles,
not to pay the large repulsion energy. This leads to a Mott insulating site. Residual kinetic energy tends to favor
antiferromagnetic exchange between nearest neighbors, to avoid Pauli principle blocking. Depending on the lattice this
can lead to various type of magnetic order. (c) If there are less than one particle per site $n<1$ but the interaction is
large, a state with only one particle per site is still favored. The conduction is only providing from the holes present in the
system that can hop at no interaction cost. Although this is still a metallic state this state has quite different properties
than the weakly correlated one described in (a).}
\end{figure}

Let us now study the properties of interacting fermions on a
lattice as described in the previous section. For
noninteracting fermions, the system remains metallic unless the
band is completely filled. The situation changes drastically in
presence of interactions. The combination of lattice and
interactions leads to a striking effect of interactions known
as the Mott transition, a phenomenon predicted
\cite{mott_historical_insulator} by Sir Nevil Mott (see
\fref{fig:mott})

To understand the phenomenon, let us consider a simple
variational calculation of the energy of the Hubbard model
(\ref{eq:hubbard}), assuming that the system is described by a
FL like ground state. We will be even more primitive and take
free electrons. In that case the ground state is simply the
Fermi sea (\ref{eq:fermi_sea}). The energy of such a state $E_0
= \bra{\psi_0} H \ket{\psi_0}$ is simply given by
\begin{equation} \label{eq:disp}
 E_0 = 2\sum_{|k|<\kF} \epsilon(k) + U \sum_i \bra{\psi_0} n_{i\up}n_{i\down} \ket{\psi_0}
\end{equation}
where the dispersion relation is (\ref{eq:tight_disp}). The
first term in (\ref{eq:disp}) is simply the average of the
kinetic energy
\begin{equation}
 \int_{-W}^{\mu_0} d\epsilon \dos(\epsilon) \epsilon = \Omega \int_{-W}^{\mu_0} d\epsilon n(\epsilon) \epsilon = \Omega \epsilon_0
\end{equation}
where $\Omega$ is the number of sites in the systems,
$n(\epsilon)$ the density of states per site, $W$ the minimum
of energy in the band and $\mu_0$ the chemical potential. For
the dispersion relation (\ref{eq:tight_disp}), $\epsilon_0$ is
a filling dependent negative number (for $n \leq 1$). The
second term in (\ref{eq:disp}) depends on the average number of
doubly occupied sites. Because the plane waves of the ground
state have a uniform amplitude on each site
\begin{equation}
 \langle n_{i\up}n_{i\down} \rangle = \langle n_{i\up} \rangle \langle n_{i\down} \rangle = \frac{n^2}{4}
\end{equation}
and thus
\begin{equation} \label{eq:disp2}
 E_0 = \Omega [\epsilon_0 + U \frac{n^2}{4}]
\end{equation}
One can notice two interesting points on this equation: i) in
the absence of interaction ($U=0$) the system gains energy by
delocalizing the particles. The maximum gain in energy, so in a
way the best metallic state, is obtained when the band is half
filled $n=1$; ii) if the system has a FL (free electron like)
ground state, the energy cost due to the interactions is
growing. Thus one could naively expect that for large $U$ a
better state could occur. Let us consider for example a
variational function in which each particle is localized on a
given site (for simplicity let us only consider $n \leq 1$)
\begin{equation}
 \ket{\psi_M} = c^\dagger_{i_1,\up}\cdots c^\dagger_{i_N,\down} \void
\end{equation}
Note that we can put the spins in an arbitrary order for the
moment. There are thus many such states $\ket{\psi_M}$
differing by their spin orientation. What is the best spin
order will be discussed in the next section. One has obviously
\begin{equation}
 \bra{\psi_M} H \ket{\psi_M} = 0
\end{equation}
since there is no kinetic energy and no repulsion given the
fact that at most one particle is on a given site. We can now
take these two states as variational estimates of what could be
the best representation of the ground state of our system, the
idea being that the wavefunction with the lower energy is the
one that has probably the largest overlap with the true ground
state of the system. At a pure variational level, this
calculation would strongly suggest that the wave function
$\ket{\psi_0}$ is a good approximation of the ground state at
small interaction while $\ket{\psi_M}$ would be the best
approximation of the ground state at large $U$. One could thus
naively expect a crossover or even a quantum phase transition
at a critical value of $U$. The nature of the two ``phases''
can be inferred from the two variational wavefunction. For
small $U$ one expects a good FL, while for large interactions
one can expect a state where the particles are localized. For
$n=1$, as shown in \fref{fig:mott}, such a state would be an
insulator. One could thus expect a metal insulator transition,
driven by the interactions at a critical value of the
interactions $U$ of the order of the kinetic energy per site
$\epsilon_0$ of the non interacting system. Of course our
little variational argument is far too primitive to allow to
seriously conclude on the above points, and one must perform
more sophisticated calculations \cite{imada_mott_review}. This
includes physical arguments, more refined variational
calculations
\cite{gutzwiller_hubbard,brinkman_correlated,yokoyama_mcv_antiferro},
slave bosons techniques \cite{kotliarnew} and refined mean
field theories \cite{kotliar_dmft_review}.

As it turns out our little calculation gives already the right
physics, originally understood by Mott. An important ingredient
is the filling of the band. Indeed the situation is extremely
different depending on whether the filling is one particle per
site or not. If the filling is one particle per site (half
filled band), then our little variational state, with one
particle per site is indeed an insulator as shown on
\fref{fig:mott}. Applying the kinetic energy operator on this
state would force it to go through a state with at least a
doubly occupied site, with an energy cost of $U$, so such
excitations could not propagate. Such an insulating state
created by interactions has been nicknamed a Mott insulator.
From the point of view of band theory and free electrons this
is a remarkable state, since a half filled system would give
normally the best type of metallic state possible. We see that
interactions are able to transform this state into an
insulating state.

If the filling is not exactly one particle per site, something
remarkable occurs again when $U > U_c$. In that case, as shown
in \fref{fig:mott}, the function were one has only singly
occupied sites is obviously still a very good starting point.
However if there are now empty sites ($n < 1$) these holes can
now propagate freely at no cost in $U$. Such a state is thus
not an insulator but still a metal. There are important
differences with the metal one would get for small $U$. Indeed
in this strongly correlated metal, only the holes can propagate
freely, and one can expect that the number of carriers will be
proportional to the doping $n-1$ and not to the total number of
particles $n$ in the system. Other properties are also
affected. I will not address in details these points but refer
the reader to the literature for more details.

To finish let me mention three important points: the first one
is that some additional phenomena can shift the Mott transition
to $U=0$ for special lattices. Indeed if the lattice is
bipartite, i.e. can be separated in two sublattices which are
connected only be the hopping operator, special properties
occur. This is in particular the case of the square or the
hexagonal lattice. On such lattices, for one particle per site
the dispersion relation has a special property known as nesting
property. Namely there is a wavevector $Q$ such that for all
$k$
\begin{equation}
 \xi(k+Q) = -\xi(k)
\end{equation}
For the tight-binding relation on a square lattice, nesting
occurs for $n=1$ and $Q=(\pi,\pi,\cdots)$. When a Fermi surface
is nested the charge and spin susceptibilities have a
logarithmic divergence due to the nesting. Typically
\begin{equation}
 \chi(Q,T) \simeq \dos(\EF) \log[\beta W]
\end{equation}
This divergent susceptibility provokes an instability of the
metallic state for arbitrary interactions. As a result, nested
systems become Mott insulators as soon as the interactions
become repulsive, and order antiferromagnetically (at least in
$d=3$ where quantum fluctuations cannot destroy the magnetic
order).

The second remark is that Mott insulators are not limited to
the case of one particle per site. Any commensurate filling can
potentially lead to an insulating state, depending on the
\emph{range} of the interactions. For example one can expect an
insulating state at $1/4$ filling if both on site and nearest
neighbor interactions are present and of sufficient strength.

The third important point is that nothing in such a mechanism
limits Mott insulators to fermionic statistics. As we saw in
the simple derivation, interactions are simply avoiding double
occupancy, which turns the system into an insulator. So we can
also expect Mott insulators to exist for bosons and even for
Bose-Fermi mixtures. The Mott phase will be essentially the
same in each case (only one particle per site), but of course
the low interaction phase will strongly depend on the precise
properties of the system (e.g. for bosons one can expect to
have a superfluid phase at small interactions).

Mott insulators are one of the most spectacular effects of
strong correlations. They are ubiquitous in nature and occur in
oxides, cuprates, organic conductors and of course are realized
in cold atomic systems both for bosons and fermions.

\subsection{Magnetic properties of Mott insulators}

\begin{figure}
\begin{center}
\includegraphics[width=0.2\linewidth]{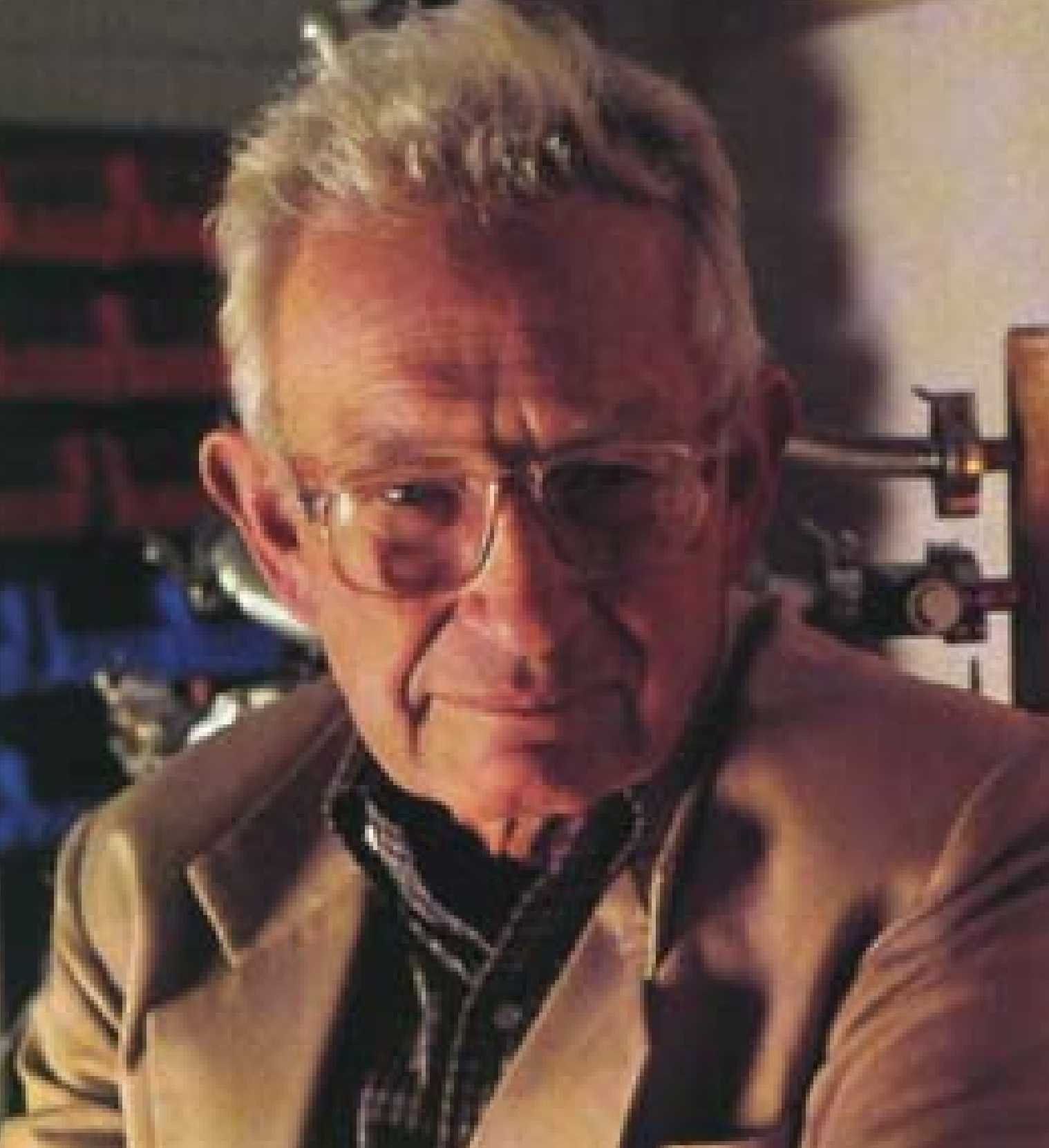}
\includegraphics[width=0.8\linewidth]{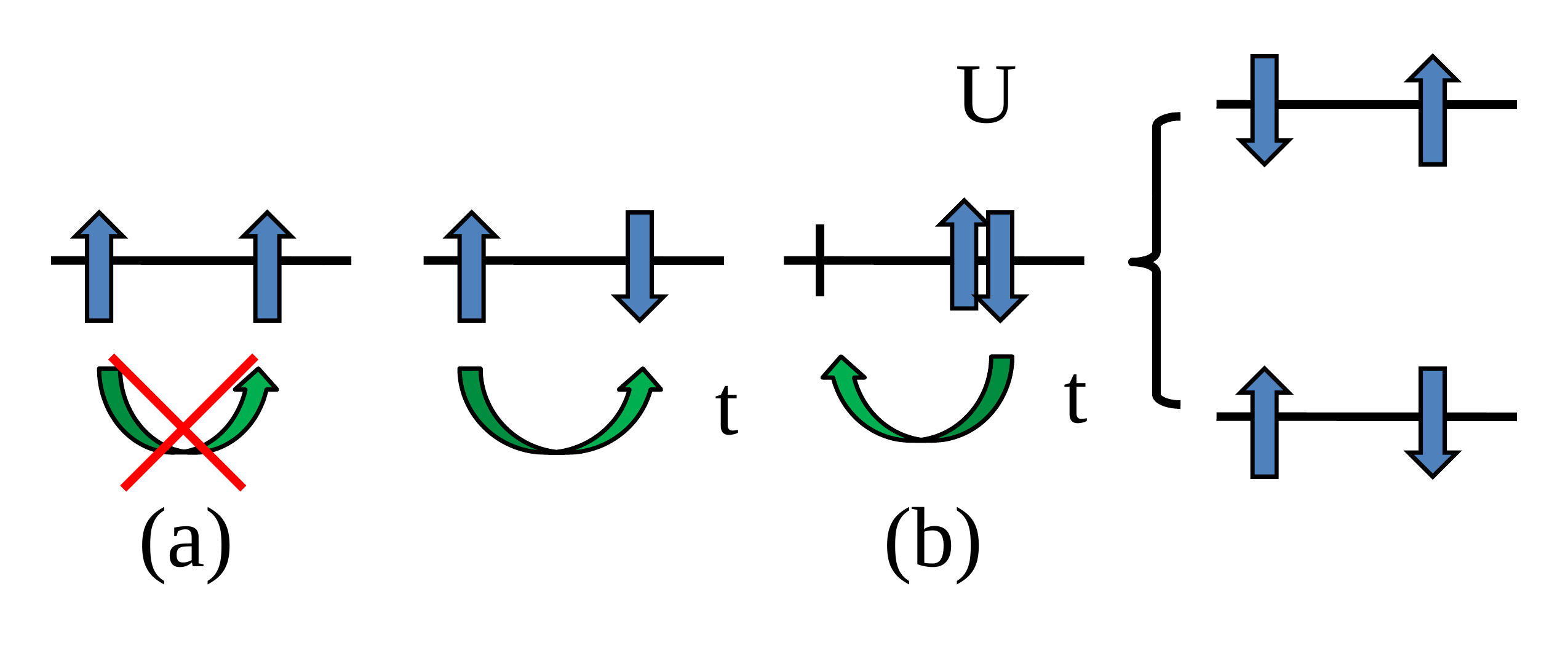}
\end{center}
\caption{\label{fig:anderson} (top) P. W. Anderson. (a) The hopping between two neighboring identical
spins is totally blocked by the Pauli principle. The system cannot gain some residual kinetic energy in
this case. (b) If the spins are antiparallel then the kinetic energy can allow virtual hopping through
an intermediate state of energy $U$. This state comes back either to the original state or to a configuration
in which the spins have been exchanged. This is the mechanism of superexchange leading to dominant antiferromagnetic
correlations for fermionic Mott insulators.}
\end{figure}

Let me concentrate now on the Mott phase ($n=1$) at large
interactions $U > U_c$. In that case, at low temperature all
charge excitations are gapped, with a gap $\Delta \sim U$ and
the system is an insulator. One could thus naively think that
one has properties very similar to the ones of a band
insulator. Although this naive expectation is, to some extent
correct for the charge properties, it is completely incorrect
for the spin ones, and very interesting spin physics occurs.
Indeed a band insulators corresponds to either zero or two
particles per site. It means that in both case the on site spin
is zero. The Pauli principle forces two particles on the same
site to be in a singlet spin state to get a fully antisymmetric
wave function. Thus a band insulator has no interesting charge
nor spin properties. This is not the case for Mott insulator
since each site is singly occupied. There is thus a spin $1/2$
degree of freedom on each site, and it is thus important to
understand the corresponding magnetic properties. This was
worked out \cite{anderson_superexchange} by P. W. Anderson (see
\fref{fig:anderson}).

To do so let us examine the case of two sites. The total
Hilbert space is
\begin{equation} \label{eq:basis}
 \ket{\up,\down}, \ket{\down,\up}, \ket{\up\down,0}, \ket{0,\up\down}, \ket{\up,\up}, \ket{\down,\down}
\end{equation}
Since the states are composed of Fermions one should be careful
with the order of operators to avoid minus signs. Let us take
the convention that
\begin{equation}
\begin{split}
 \ket{\up,\down} = \hc{c}_{1\up} \hc{c}_{2\down} \void \quad &, \quad
 \ket{\down\up} = \hc{c}_{1\down} \hc{c}_{2\up} \void \\
  \ket{\up\down,0} = \hc{c}_{1\up} \hc{c}_{1\down} \void \quad &, \quad
 \ket{0,\up\down} = \hc{c}_{2\up} \hc{c}_{2\down} \void
\end{split}
\end{equation}
The states with two particles per site are states of energy
$\sim U$ and therefore strongly suppressed. We thus need to
find what is the form of the Hamiltonian when restricted to the
states with only one particle per site. It is easy to check
that the two states $\ket{\up\up}$ and $\ket{\down\down}$ are
eigenstates of $H$
\begin{equation}
 H \ket{\up\up} = 0
\end{equation}
and a similar equation for $\ket{\down\down}$. The reason why
the kinetic energy does not act on such a state is shown in
\fref{fig:anderson}. The Pauli principle block the hopping if
the two spins are equal. On the contrary, if the two spins are
opposite the particles can make a virtual jump on the
neighboring site. Since this state is of high energy $U$ the
particles must come back to the original position, or the two
particles can exchange leading to a term similar to a spin
exchange (see \fref{fig:anderson}). The \emph{kinetic energy}
thus leads to a \emph{magnetic} exchange, named superexchange,
that will clearly favor configurations with opposite
neighboring spins, namely will be antiferromagnetic.

Let us now quantify the mechanism. The Hamiltonian can be
written in the basis (\ref{eq:basis}) (only the action on the
first four states is shown since $\ket{\up\up}$ and
$\ket{\down\down}$ are eigenstates and thus are uncoupled to
the other ones)
\begin{equation}
H = \left(
\begin{array}{cccc}
  0 & 0 & -t & -t \\
  0 & 0 &  t &  t \\
 -t & t &  U &  0 \\
 -t & t &  0 &  U
\end{array}
\right)
\end{equation}
This Hamiltonian couples the low energy states with one
particle per site to high energy states of energy $U$. In order
to find the restriction of $H$ to the low energy sector, let us
first make a canonical transformation of $H$
\begin{equation}
 H' = e^{i S} H e^{-i S} \simeq H + i [S,H] + \frac{i^2}2 [S,[S,H]] + \cdots
\end{equation}
where the matrix $S$ is expected to be perturbative in $t/U$.
In this transformation one wants to chose the matrix $S$ such
that
\begin{equation} \label{eq:condition}
 H_t + i [S, H_U] = 0
\end{equation}
This will ensure that $H'$ has no elements connecting the low
energy sector with the sector of energy $U$. The restriction of
$H'$ to the low energy sector will thus be the Hamiltonian we
need to diagonalize to find the spin properties.

Using the condition (\ref{eq:condition}) and keeping only terms
up to order $t^2/U$ it is easy to check that
\begin{equation} \label{eq:canonic}
 H' = H_U + \frac{i}2 [S, H_t]
\end{equation}
Since $H$ is block diagonal one can easily determine $S$ to be
\begin{equation}
S = \frac{i}U \left(
\begin{array}{cccc}
  0 & 0 & -t & -t \\
  0 & 0 &  t &  t \\
 t & -t &  0 &  0 \\
 t & -t &  0 &  0
\end{array}
\right)
\end{equation}
This leads using (\ref{eq:canonic}) to
\begin{equation}
 H' = \frac{4 t^2}U [\frac12 [\ket{\up\down}\bra{\down\up} + \ket{\down\up}\bra{\up\down}] - \frac12[\ket{\up\down}\bra{\up\down} + \ket{\down\up}\bra{\down\up}]
\end{equation}
This Hamiltonian must be complemented by the fact that $H'
\ket{\up\up} = 0$. Since there is one particle per site, it can
now be represented by a spin $1/2$ operator $S^\alpha = \frac12
\sigma^\alpha$, where $\alpha=x,y,z$ and the $\sigma$ are the
Pauli matrices. Using $S^+ = S^x + i S^y$ for the raising spin
operator one obtains
\begin{equation}
  H' = \frac{4 t^2}U [\frac12 [S^+_1 S^-_2 + S^-_1 S^+_2] + S^z_1 S^z_2  - \frac14]
\end{equation}
Up to a constant which is a simple shift of the origin of
energies, the final Hamiltonian is thus the Heisenberg
Hamiltonian
\begin{equation} \label{eq:heisenberg}
 H = J \sum_{\langle i j \rangle} \vS_i \cdot \vS_j
\end{equation}
where the magnetic exchange is $J = 4t^2/U$.

As we saw for fermions, the exchange $J$ is positive and thus
the system has dominant antiferromagnetic correlations. This
explain naturally the remarkable fact that antiferromagnetism
is quite common in solid state physics. Of course the magnetic
properties depend on the lattice, and the Hamiltonian
(\ref{eq:heisenberg}) can potentially lead to very rich physics
\cite{auerbach_book_spins}. It is important to note that we are
dealing here with quantum spins for which $[S_x,S_y] = i S_z$
and thus the three component of the spins cannot be determined
simultaneously. Quantum fluctuation will thus drastically
affect the possible spin orders. Depending on the lattice
various ground states are possible ranging from spin liquids to
ordered states.

\subsection{Summary of the basic models}

Let me summarize in this section the basic models that are
commonly used to tackle the properties of strongly correlated
systems. Of course these are just the simplest possible case,
and many extensions and interesting generalization have been
proposed. There is currently a great interest in realizing
these models with cold atomic gases.

\subsubsection{Fermions}

The simplest interacting model on a lattice is the Hubbard
model
\begin{equation} \label{eq:hubbcan}
 H = -t \sum_{\langle ij \rangle, \sigma} \hc{c}_{i\sigma} c_{i\sigma} + U \sum_i n_{i\up}n_{i\down}
\end{equation}
where $\langle\rangle$ denote the nearest neighbors, which of
course depends on the topology of the lattice (square,
triangular, etc.). Having only on site interactions this model
can only stabilize a Mott insulating phase at half filling
($n=1$). If the lattice is bipartite (e.g. square), the Mott
phase is stable for any $U >0$ and the ground state is an
antiferromagnetic state. If the lattice is not bipartite, then
a critical value of $U$ is normally requested to reach the Mott
insulating state. The spin order then depends on the lattice
(see the spin section below).

For $U<0$ this model has a BCS type instability leading to a
superconducting ground state. There are interesting symmetries
for the Hubbard model. In particular for bipartite lattices
there is a particle-hole symmetry for one spin species that can
map the repulsive Hubbard model onto the attractive one, and
exchange the magnetic field and the chemical potential. This
symmetry can be exploited in several context, but in particular
can be very useful in the cold atom context for probing the
phases of the repulsive Hubbard model in a much more convenient
way \cite{ho_attractive_hubbard}.

The canonical Hubbard model can be easily extended in several
ways. Longer range hopping can be added, and of course longer
range interactions than on-site can also be included. In that
case insulating phases can in general be stabilized at other
commensurate fillings. Finally one can also increase the number
of states per site (so called multi-orbital Hubbard model), a
situation that has been useful in condensed matter and also is
potentially relevant for cold atoms in optical lattices, for
example if one needs to take into account the higher states in
each well of \fref{fig:optical}.

\subsubsection{Bosons}

Here also the canonical model is the Hubbard model (nicknamed
Bose-Hubbard model in that case). The main difference is that
the simplest case do not need spins for the bosons. The
simplest Bose-Hubbard model is thus
\begin{equation}
 H_B = -t \sum_{\langle ij \rangle} [\hc{b}_i b_j + \hc{b}_j b_i] + U \sum_i n_i(n_i - 1)
\end{equation}
This model has also in general a Mott transition for
sufficiently large interaction when the filling is commensurate
$n=1$ (1 boson per site)
\cite{haldane_bosons,fisher_boson_loc}. In a similar way than
for the fermions the model can be extended to the case of
longer range hopping or longer range interactions, allowing for
more insulating phases at other commensurabilities than for
$n=1$.

The case of the Hubbard model with a nearest neighbor
interaction
\begin{equation}
 U \sum_i n_i(n_i-1) + V \sum_{\langle i j\rangle} n_i n_j
\end{equation}
has an interesting property that is worth noting. When $U$ and
$V$ are of the same order of magnitude, the system can have
fluctuations of charge on a site going between zero, one and
two bosons per site, since putting two bosons per site is a way
to escape paying the repulsion $V$. The system is thus very
close to a system with three state per site, i.e. of a system
that can be mapped onto a spin $1$. In one dimension spin $1$
are known to have very special properties, and thus similar
properties are expected for such an extended Hubbard model
\cite{berg_haldane_cold_bosons}. In particular they can have a
topologically ordered phase (Haldane phase \cite{haldane_gap}
for spin one). In higher dimensions such models can have
complex orders, and there is in particular a debate on whether
such models can sustain simultaneously a crystalline order
(Mott phase for the bosons) and superfluidity, the so called
supersolid phase \cite{NiyazBatrouni1994,wessel_supersolid}.

Various extensions of the canonical Bose-Hubbard are worth
noting. The most natural one, in connection with cold atomic
systems is to consider a model with more than one species, in
other word re-introducing a kind of ``spin for the bosons''.
This will correspond to bosonic mixtures. In that case one can
generally expect interactions of the form
\begin{equation}
 H = U_{\up\up} n_\up (n_\up -1) + U_{\down\down} n_\down (n_\down -1) + U_{\up\down} n_\up n_\down
\end{equation}
Note that the $U_{\sigma\sigma}$ interactions did not exist for
the fermionic Hubbard model because of the Pauli principle. For
bosons their presence allow for a very rich physics. In
particular the nature of the superexchange interactions will
depend on the difference \cite{DuanLukin2003} between the inter
and the intra species interactions. If $U_{\up\up},
U_{\down\down} \gg U_{\up\down}$ one is in a situation very
similar to the case of fermions, with a dominant
antiferromagnetic exchange. The ground state of such a model
will be antiferromagnetic. On the contrary, if one is in the
opposite situation then it is more favorable for the kinetic
energy to have parallel spins nearby and the superexchange will
be ferromagnetic. This leads to the possibility of new ground
states, and even new phases in one dimension
\cite{zvonarev_ferro_cold}.

\subsubsection{Spin systems}

If the charge degrees of freedom are localized, usually but not
necessarily for one particle per site, the resulting
Hamiltonian only involves the spins degrees of freedom. The
simplest one is for fermions with spin 1/2, and is the
Heisenberg Hamiltonian
\begin{equation}  \label{eq:heisenberg2}
 H = J \sum_{\langle ij \rangle} \vS_i \cdot \vS_j
\end{equation}
Depending on the spin exchange a host of magnetic phase can
exist. As discussed above Fermions lead mostly to
antiferromagnetic exchanges, while bosons with two degrees of
freedom would mostly be ferromagnetic.

Let me mention a final mapping which is quite useful in
connecting the spins and itinerant materials. Spin $1/2$ can in
fact be mapped back onto bosons. The general transformation has
been worked out by Holstein and Primakov for a spin $S$, but
let me specialize here to the case of spin $1/2$, which has
been worked out by Matsubara and Matsuda
\cite{matsubara_spins_bosons_mapping} and is particularly
transparent. Since the Hilbert space of spin $1/2$ has only two
states, they can be mapped onto the presence and absence of a
boson by the mapping
\begin{equation}\label{eq:mapbos}
\begin{split}
 S^+ &= \hc{b} \\
 S^z &= n_i - 1/2 = \hc{b}_i b_i - 1/2
\end{split}
\end{equation}
because only two states are possible for the spins, it is
important to put a hard core constraint on the bosons, imposing
that at most one boson can exist on a given site.

The Hamiltonian (\ref{eq:heisenberg2}), where we have
introduced two coupling constants $J_{XY}$ and $J_Z$ for the
two corresponding terms, can be rewritten as
\begin{equation}
 H = \frac{J_{XY}}2 \sum_{\langle ij \rangle} [S^+_i S^-_j + S^-_i S^+_j] + J_Z \sum_{\langle ij \rangle} S^z_i S^z_j
\end{equation}
which becomes using the mapping (\ref{eq:mapbos})
\begin{equation}
 H = \frac{J_{XY}}2 \sum_{\langle ij \rangle} [b^\dagger_i b_j + b_i b^\dagger_j] + J_Z \sum_{\langle ij \rangle} (n_i-1/2) (n_j-1/2)
\end{equation}
The hard core constraint can be imposed by adding an on site
interaction $U$ and letting it go to infinity. It is thus
possible to map a spin model onto
\begin{equation}
 H = \frac{J_{XY}}2 \sum_{\langle ij \rangle} [b^\dagger_i b_j + b_i b^\dagger_j] + U \sum_i n_i (n_i - 1) +
 J_Z \sum_{\langle ij \rangle} (n_i-1/2) (n_j-1/2)
\end{equation}
in the limit $U \to \infty$. One recognize the form of an
extended Hubbard model.

This mapping allows not only to solve certain bosonic problems
by borrowing the intuition on magnetic order or vice versa, but
it also allows to use spin systems to experimentally realize
Bose-Einstein condensate and study their critical properties.
This has been a line of investigation that has been very
fruitfully pursued recently \cite{giamarchi_BEC_dimers_review}.

\section{One dimensional systems} \label{sec:onedim}

Let us now turn to one dimensional systems. As we discussed in
the previous sections the effect of interactions is maximal
there. For fermions this leads to the destruction of the Fermi
liquid state. For bosons, it is easy to see that simple BEC
states are likewise impossible. Due to quantum fluctuations it
is impossible to break a continuous symmetry (here the phase
symmetry of the wave function) in one dimension. One has thus
to face a radically different physics than for their higher
dimensional counterparts. Fortunately the one-dimensional
character brings new physics but also new methods to tackle the
problem. This allows for quite complete solutions to be
obtained, revealing remarkable physics phenomena and challenges
\cite{giamarchi_book_1d}.

I will not cover here all these developments and physical
realizations. I have written a whole book on the subject
\cite{giamarchi_book_1d} where the interested reader can find
this information in a much more detailed and pedagogical way
than the size of these note allow to present. I will however
present the very basic ideas here.

Before we embark on the description of the physics, a short
historical note. To solve one dimensional systems, crucial
theoretical progress were made, mostly in the 1970's allowing a
detailed understanding of the properties of such systems. This
culminated in the 1980's with a new concept of interacting
one-dimensional particles, analogous to the Fermi liquid for
interacting electrons in three dimensions: the Luttinger liquid
\cite{haldane_bosonisation,haldane_effective_harmonic_fluid_approach}.
Since then many developments have enriched further our
understanding of such systems \cite{giamarchi_book_1d}, ranging
from conformal field theory to important progress in the exact
solutions such as Bethe ansatz. In addition to these important
theoretical progress, experimental realizations have knows
comparably spectacular developments. One-dimensional systems
were initially a theorist's toy. Experimental realizations
started to appear in the 1970's with polymers and organic
compounds. But in the last 20 years or so we have seen a real
explosion of realization of one-dimensional systems. The
progress in material research made it possible to realize bulk
materials with one-dimensional structures inside. The most
famous ones are the organic superconductors
\cite{lebed_book_1d} and the spin and ladder compounds
\cite{dagotto_ladder_review}. At the same time, the tremendous
progress in nanotechnology allowed to obtain realizations of
isolated one-dimensional systems such as quantum wires
\cite{fisher_transport_luttinger_review}, Josephson junction
arrays \cite{fazio_josephson_junction_review}, edge states in
quantum hall systems \cite{wen_edge_review}, and nanotubes
\cite{dresselhaus_book_fullerenes_nanotubes}. Last but not
least, the recent progress in Bose condensation in optical
traps have allowed an unprecedented way to probe for strong
interaction effects in such systems
\cite{pitaevskii_becbook,greiner_mott_bec,bloch_cold_atoms_optical_lattices_review}.

\subsection{Realization of one dimensional systems}
\label{sec:1dcold}

Let us first discuss how one can obtain ``one dimensional''
objects in a real three dimensional world. All the one
dimensional systems are characterized by a confining potential
forcing the particles to be in a localized state.
\begin{figure}
\begin{center}
  \includegraphics[width=0.8\linewidth]{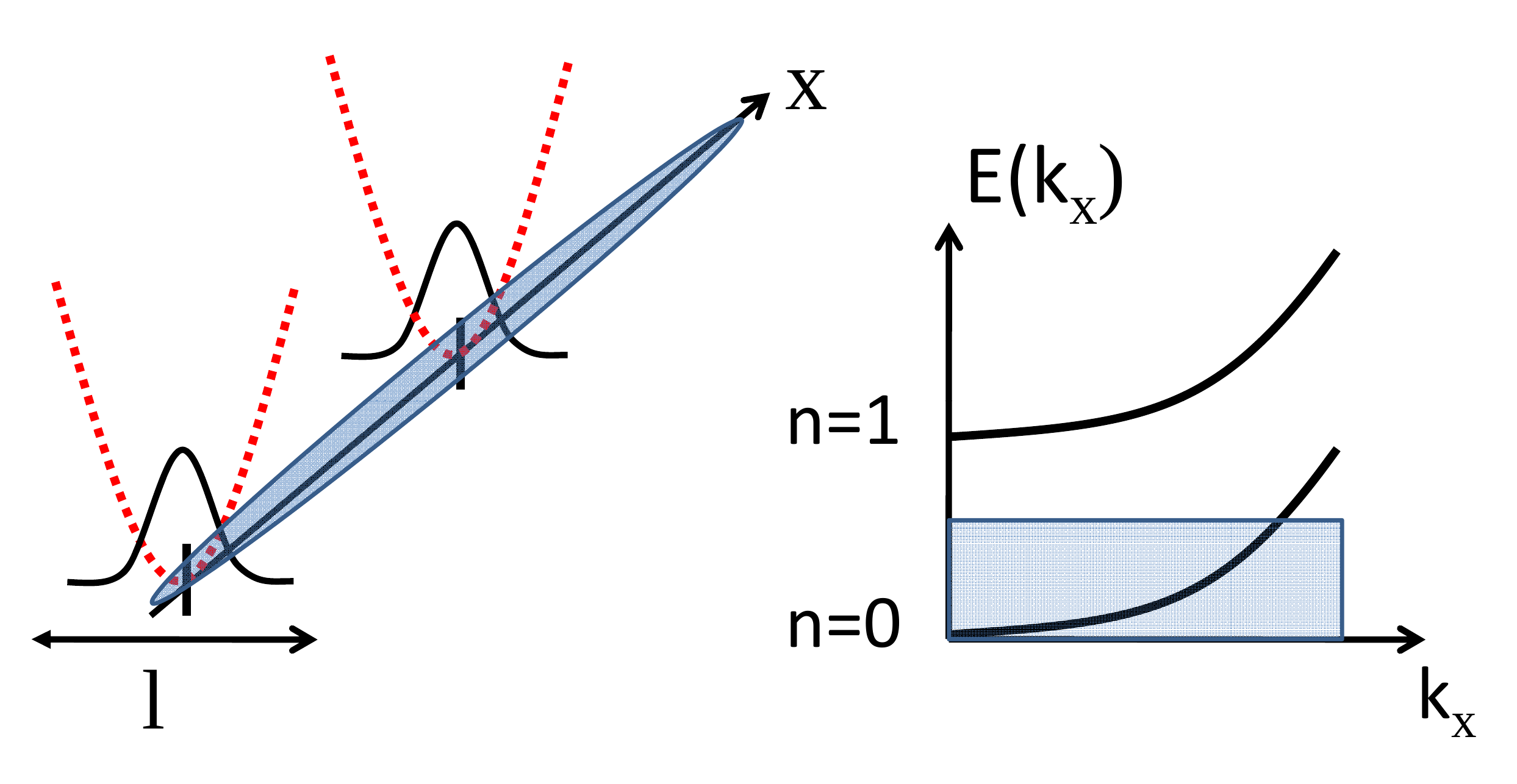}
\end{center}
 \caption{\label{fig:confwire} (left) Confinement of the electron gas in a one-dimensional tube of transverse size $l$.
 $x$ is the direction of the tube. Only one transverse direction of
 confinement has been shown for clarity. Due to the transverse confining
 potential the transverse degrees of freedom are strongly quantized.
 (right) Dispersion relation $E(k)$. Only half of the dispersion relation is shown for clarity.
 $k$ is the momentum parallel to the tube direction.
 The degrees of freedom transverse to the tube direction lead to the formation of minibands,
 labelled by a quantum number $n$. If only one miniband is populated,
 as represented by the gray box, the system is equivalent
 to a one dimensional system where only longitudinal degrees of
 freedom can vary.}
\end{figure}
The wavefunction of the system is thus of the form
\begin{equation}
 \psi(x,r_\perp) = e^{i k x} \phi(r_\perp)
\end{equation}
where $\phi$ depends on the precise form of the confining
potential For an infinite well, as show in \fref{fig:confwire},
$\phi$ is $\phi(y) = \sin((2n_y+1)\pi y/l)$, whereas it would
be a gaussian function (\ref{eq:harmfond}) for an harmonic
confinement. The energy is of the form
\begin{equation}
 E = \frac{k^2}{2 m} + \frac{k_y^2}{2 m}
\end{equation}
where for simplicity I have taken hard walls confinement. Due
to the narrowness of the transverse channel $l$, the
quantization of $k_y$ is sizeable. Indeed, the change in energy
by changing the transverse quantum number $n_y$ is at least
(e.g. $n_y =0$ to $n_y=1$)
\begin{equation}
 \Delta E = \frac{3\pi^2}{2 m l^2}
\end{equation}
This leads to minibands as shown in \fref{fig:confwire}. If the
distance between the minibands is larger than the temperature
or interactions energy one is in a situation where only one
miniband can be excited. The transverse degrees of freedom are
thus frozen and only $k$ matters. The system is a
one-dimensional quantum system.

\subsection{Bosonization dictionary} \label{sec:bozo}

Treating interacting particle in one dimension is a quite
difficult task. One very interesting technique is provided by
the so-called bosonization. It has the advantage of giving a
very simple description of the low energy properties of the
system, and of being completely general and very useful for
many one dimensional systems. This chapter describe its vary
basic features. For more details and physical insights on this
technique both for fermions and bosons I refer the reader to
\cite{giamarchi_book_1d}.

The idea behind the bosonization technique is to reexpress the
excitations of the system in a basis of collective excitations.
Indeed in one dimension it is easy to realize that single
particle excitations cannot really exit. One particle when
moving will push its neighbors and so on, which means that any
individual motion is converted into a collective one.
Collective excitations should thus be a good basis to represent
a one dimensional system.

\begin{figure}
\begin{center}
  \includegraphics[width=0.8\linewidth]{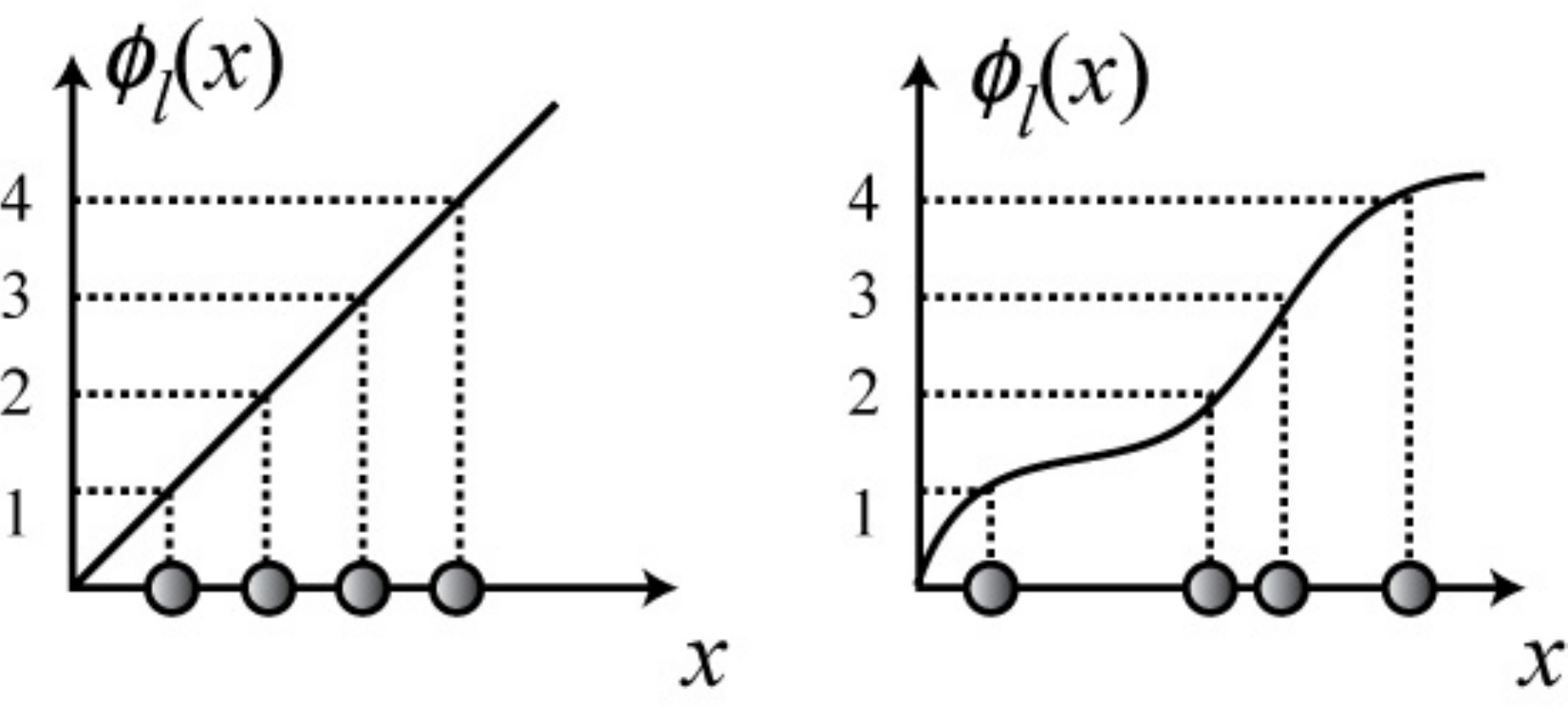}
\end{center}
 \caption{Some examples of the labelling field $\phi_l(x)$. If the
 particles form a perfect lattice of lattice spacing $d$, then
 $\phi_l^0(x) = 2\pi x/d$, and is just a straight line.
 Different functions
 $\phi_l(x)$ allow to put the particles at any position in space. Note that $\phi(x)$ is always
 an increasing function regardless of the position of the particles. [From \protect\cite{giamarchi_book_1d}]}
 \label{fig:labelfield}
\end{figure}

To exploit this idea, let us start with the density operator
\begin{equation}\label{eq:densmoche}
 \rho(x) = \sum_i \delta(x-x_i)
\end{equation}
where $x_i$ is the position operator of the $i$th particle. We
label the position of the $i$th particle by an `equilibrium'
position $R_i^0$ that the particle would occupy if the
particles were forming a perfect crystalline lattice, and the
displacement $u_i$ relative to this equilibrium position. Thus,
\begin{equation}
 x_i = R_i^0 + u_i
\end{equation}
If $\rho_0$ is the average density of particles,
$d=\rho_0^{-1}$ is the distance between the particles. Then,
the equilibrium position of the $i$th particle is
\begin{equation}
 R_i^0 = d i
\end{equation}
Note that at that stage it is not important whether we are
dealing with fermions or bosons. The density operator written
as (\ref{eq:densmoche}) is not very convenient. To rewrite it
in a more pleasant form we introduce a labelling field
$\phi_l(x)$ \cite{haldane_bosons}. This field, which is a
continuous function of the position, takes the value
$\phi_l(x_i) = 2\pi i$ at the position of the $i$th particle.
It can thus be viewed as a way to number the particles. Since
in one dimension, contrary to higher dimensions, one can always
number the particles in an unique way (e.g. starting at
$x=-\infty$ and processing from left to right), this field is
always well-defined. Some examples are shown in
\fref{fig:labelfield}.
Using this labelling field and the rules for transforming
$\delta$ functions
\begin{equation}
 \delta(f(x)) = \sum_{\mbox{zeros of $f$}} \frac1{|f'(x_i)|}
 \delta(x-x_i)
\end{equation}
one can rewrite the density as
\begin{eqnarray}
 \rho(x) &=& \sum_i \delta(x-x_i) \nonumber \\
  &=& \sum_n |\nabla \phi_l(x)| \delta(\phi_l(x) - 2\pi n)
  \label{eq:denslab}
\end{eqnarray}
It is easy to see from \fref{fig:labelfield} that $\phi_l(x)$
can always be taken as an increasing function of $x$, which
allows to drop the absolute value in (\ref{eq:denslab}). Using
the Poisson summation formula this can be rewritten
\begin{equation}
 \rho(x) = \frac{\nabla \phi_l(x)}{2\pi} \sum_p e^{i p \phi_l(x)}
\end{equation}
where $p$ is an integer. It is convenient to define a field
$\phi$ relative to the perfect crystalline solution and to
introduce
\begin{equation}
 \phi_l(x) = 2 \pi \rho_0 x - 2\phi(x)
\end{equation}
The density becomes
\begin{equation} \label{eq:locintbos}
 \rho(x) = \left[\rho_0 -\frac{1}{\pi} \nabla \phi(x)\right] \sum_p e^{i 2 p (\pi \rho_0 x - \phi(x))}
\end{equation}
Since the density operators at two different sites commute it
is normal to expect that the field $\phi(x)$ commutes with
itself. Note that if one averages the density over distances
large compared to the interparticle distance $d$ all
oscillating terms in (\ref{eq:locintbos}) vanish. Thus, only
$p=0$ remains and this smeared density is
\begin{equation} \label{eq:smeardens}
 \rho_{q\sim 0}(x) \simeq \rho_0 - \frac1\pi \nabla\phi(x)
\end{equation}
We can now write the single-particle creation operator
$\psi^\dagger(x)$. Such an operator can always be written as
\begin{equation} \label{eq:singlephen}
 \psi^\dagger(x) = [\rho(x)]^{1/2}e^{-i \theta(x)}
\end{equation}
where $\theta(x)$ is some operator. In the case where one would
have Bose condensation, $\theta$ would just be the superfluid
phase of the system. The commutation relations between the
$\psi$ impose some commutation relations between the density
operators and the $\theta(x)$. For bosons, the condition is
\begin{equation} \label{eq:comphen}
 [\psi^{\phantom\dagger}_B(x),\psi_B^\dagger(x')] = \delta(x-x')
\end{equation}
Using (\ref{eq:singlephen}) the commutator gives
\begin{equation} \label{eq:comexpl}
 e^{+i \theta(x)}[\rho(x)]^{1/2}[\rho(x')]^{1/2}e^{-i \theta(x')}
 -[\rho(x')]^{1/2}e^{-i \theta(x')}e^{+i \theta(x)}[\rho(x)]^{1/2}
\end{equation}
If we assume quite reasonably that the field $\theta$ commutes
with itself ($[\theta(x),\theta(x')] = 0$), the commutator
(\ref{eq:comexpl}) is obviously zero for $x \ne x'$ if (for
$x\ne x'$)
\begin{equation} \label{eq:comcompl}
 [[\rho(x)]^{1/2},e^{-i \theta(x')}] = 0
\end{equation}
A sufficient condition to satisfy (\ref{eq:comphen}) would thus
be
\begin{equation}\label{eq:commutbas}
 [\rho(x),e^{-i\theta(x')}] = \delta(x-x')e^{-i\theta(x')}
\end{equation}
It is easy to check that if the density were only the smeared
density (\ref{eq:smeardens}) then (\ref{eq:commutbas}) is
obviously satisfied if
\begin{equation} \label{eq:conjphi}
 [\frac1\pi \nabla\phi(x),\theta(x')] = -i \delta(x-x')
\end{equation}
One can show that this is indeed the correct condition to use
\cite{giamarchi_book_1d}. Equation (\ref{eq:conjphi}) proves
that $\theta$ and $\frac1\pi \nabla\phi$ are canonically
conjugate. Note that for the moment this results from totally
general considerations and does not rest on a given microscopic
model. Such commutation relations are also physically very
reasonable since they encode the well known duality relation
between the superfluid phase and the total number of particles.
Integrating by part (\ref{eq:conjphi}) shows that
\begin{equation}
 \pi\Pi(x) = \hbar\nabla\theta(x)
\end{equation}
where $\Pi(x)$ is the canonically conjugate momentum to
$\phi(x)$.

To obtain the single-particle operator one can substitute
(\ref{eq:locintbos}) into (\ref{eq:singlephen}). Since the
square root of a delta function is also a delta function up to
a normalization factor the square root of $\rho$ is identical
to $\rho$ up to a normalization factor that depends on the
ultraviolet structure of the theory. Thus,
\begin{equation} \label{eq:singlebos}
 \psi^\dagger_B(x) = [\rho_0 - \frac1\pi\nabla \phi(x)]^{1/2}
 \sum_{p} e^{i 2 p (\pi \rho_0 x - \phi(x))}e^{-i \theta(x)}
\end{equation}
where the index $B$ emphasizes that this is the representation
of a \emph{bosonic} creation operator.

How to modify the above formulas if we have fermions instead of
bosons? The density can obviously be expressed in the same way
in terms of the field $\phi$. For the single-particle operator
one has to satisfy an anticommutation relation instead of
(\ref{eq:comphen}). We thus have to introduce in representation
(\ref{eq:singlephen}) something that introduces the proper
minus sign when the two fermions operators are commuted. This
is known as a Jordan--Wigner transformation. Here, the operator
to add is easy to guess. Since the field $\phi_l$ has been
constructed to be a multiple of $2\pi$ at each particle,
$e^{i\frac12\phi_l(x)}$ oscillates between $\pm 1$ at the
location of consecutive particles. The Fermi field can thus be
easily constructed from the boson field (\ref{eq:singlephen})
by
\begin{equation}
    \psi^\dagger_F(x) = \psi^\dagger_B(x) e^{i\frac12\phi_l(x)}
\end{equation}
This can be rewritten in a form similar to
(\ref{eq:singlephen}) as
\begin{equation} \label{eq:singlephenfer}
 \psi^\dagger_F(x) = [\rho_0 - \frac1\pi \nabla \phi(x)]^{1/2}
 \sum_{p} e^{i (2p+1) (\pi \rho_0 x - \phi(x))}e^{-i \theta(x)}
\end{equation}
The above formulas are a way to represent the excitations of
the system directly in terms of variables defined in the
continuum limit.

The fact that all operators are now expressed in terms of
variables describing \emph{collective} excitations is at the
heart of the use of such representation, since as already
pointed out, in one dimension excitations are necessarily
collective as soon as interactions are present. In addition the
fields $\phi$ and $\theta$ have a very simple physical
interpretation. If one forgets their canonical commutation
relations, order in $\theta$ indicates that the system has a
coherent phase as indicated by (\ref{eq:singlebos}), which is
the signature of superfluidity. On the other hand order in
$\phi$ means that the density is a perfectly periodic pattern
as can be seen from (\ref{eq:locintbos}). This means that the
system has ``crystallized''. For fermions note that the least
oscillating term in (\ref{eq:singlephenfer}) corresponds to
$p=\pm 1$. This leads to two terms oscillating with a period
$\pm\pi\rho_0$ which is nothing but $\pm\kF$. These two terms
thus represent the Fermions leaving around their respective
Fermi points $\pm\kF$, also known as right movers and left
movers.

\subsection{Physical results and Luttinger liquid}

To determine the Hamiltonian in the bosonization representation
we use (\ref{eq:singlebos}) in the kinetic energy of bosons. It
becomes
\begin{equation}
 H_K \simeq \int dx \frac{\hbar^2\rho_0}{2m}(\nabla e^{i\theta})(\nabla e^{-i\theta})
     = \int dx \frac{\hbar^2\rho_0}{2m} (\nabla\theta)^2
\end{equation}
which is the part coming from the single-particle operator
containing less powers of $\nabla\phi$ and thus the most
relevant. Using (\ref{eq:boscont}) and (\ref{eq:locintbos}),
the interaction term becomes
\begin{equation} \label{eq:intbosbos}
 H_{\rm int} = \int dx V_0 \frac{1}{2\pi^2} (\nabla\phi)^2
\end{equation}
plus higher order operators. Keeping only the above lowest
order shows that the Hamiltonian of the interacting bosonic
system can be rewritten as
\begin{equation} \label{eq:luthamphen}
 H = \frac{\hbar}{2\pi}\int dx [\frac{u K}{\hbar^2} (\pi \Pi(x))^2 + \frac{u}{K}
 (\nabla\phi(x))^2]
\end{equation}
where I have put back the $\hbar$ for completeness. This leads
to the action
\begin{equation} \label{eq:lutacphen}
 S/\hbar = \frac1{2\pi K} \int dx \;d\tau [\frac1u (\partial_\tau\phi)^2 +
 u (\partial_x\phi(x))^2]
\end{equation}
This hamiltonian is a standard sound wave one. The fluctuation
of the phase $\phi$ represent the ``phonon'' modes of the
density wave as given by (\ref{eq:locintbos}). One immediately
sees that this action leads to a dispersion relation, $\omega^2
= u^2 k^2$, i.e. to a linear spectrum. $u$ is the velocity of
the excitations. $K$ is a dimensionless parameter whose role
will be apparent below.  The parameters $u$ and $K$ are used to
parameterize the two coefficients in front of the two
operators. In the above expressions they are given by
\begin{equation} \label{eq:ukpert}
\begin{split}
 u K & = \frac{\pi  \hbar \rho_0}m \\
 \frac{u}{K} &= \frac{V_0}{\hbar \pi }
\end{split}
\end{equation}
This shows that for weak interactions $u \propto (\rho_0
V_0)^{1/2}$ while $K \propto (\rho_0 / V_0)^{1/2}$. In
establishing the above expressions we have thrown away the
higher order operators, that are less relevant. The important
point is that these higher order terms will not change the form
of the Hamiltonian (like making cross terms between $\phi$ and
$\theta$ appears etc.) but {\it only} renormalize the
coefficients $u$ and $K$ (for more details see
\cite{giamarchi_book_1d}). For fermions it is easy to check
that one obtains a similar form. The important difference is
that since the single particle operator contains already $\phi$
and $\theta$ at the lowest order (see (\ref{eq:singlephenfer}))
the kinetic energy alone leads to $K=1$ and interactions
perturb around this value, while for bosons non-interacting
bosons correspond to $K = \infty$.

The low-energy properties of interacting quantum fluids are
thus described by an Hamiltonian of the form
(\ref{eq:luthamphen}) \emph{provided} the proper $u$ and $K$
are used. These two coefficients \emph{totally} characterize
the low-energy properties of massless one-dimensional systems.
The bosonic representation and Hamiltonian
(\ref{eq:luthamphen}) play the same role for one-dimensional
systems than the Fermi liquid theory plays for
higher-dimensional systems. It is an effective low-energy
theory that is the fixed point of any massless phase,
regardless of the precise form of the microscopic Hamiltonian.
This theory, which is known as Luttinger liquid theory
\cite{haldane_bosonisation,haldane_bosons}, depends only on the
two parameters $u$ and $K$. Provided that the correct value of
these parameters are used, \emph{all} asymptotic properties of
the correlation functions of the system then can be obtained
\emph{exactly} using (\ref{eq:locintbos}) and
(\ref{eq:singlebos}) or (\ref{eq:singlephenfer}).

Computing the Luttinger liquid coefficient can be done very
efficiently. For small interaction, perturbation theory such as
(\ref{eq:ukpert}) can be used. More generally one just needs
two relations involving these coefficients to obtain them.
These could be for example two thermodynamic quantities, which
makes it easy to extract from either Bethe-ansatz solutions if
the model is integrable or numerical solutions. The Luttinger
liquid theory thus provides, coupled with the numerics, an
incredibly accurate way to compute correlations and physical
properties of a system (see e.g.
\cite{klanjsek_nmr_ladder_luttinger} for a remarkable example).
For more details on the various procedures and models see
\cite{giamarchi_book_1d}. But, of course, the most important
use of Luttinger liquid theory is to justify the use of the
boson Hamiltonian and fermion--boson relations as starting
points for any microscopic model. The Luttinger parameters then
become some effective parameters. They can be taken as input,
based on general rules (e.g. for bosons $K=\infty$ for non
interacting bosons and $K$ decreases as the repulsion
increases, for other general rules see
\cite{giamarchi_book_1d}), without any reference to a
particular microscopic model. This removes part of the
caricatural aspects of any modelization of a true experimental
system. This use of the Luttinger liquid is reminiscent of the
one where perturbations (impurity, electron-phonon interactions
etc.) are added on the Fermi liquid theory. The calculations in
$d=1$ proceed in the same spirit with the Luttinger liquid
replacing the Fermi liquid. The Luttinger liquid theory is thus
an invaluable tool to tackle the effect of perturbations on an
interacting one-dimensional electron gas (such as the effect of
lattice, impurities, coupling between chains, etc.). I refer
the reader to \cite{giamarchi_book_1d} for more on those
points.

\subsection{Correlations}

Let us now examine in details the physical properties of such a
Luttinger liquid. For this we need the correlation functions. I
briefly show here how to compute them using the standard
operator technique. More detailed calculations and functional
integral methods are given in \cite{giamarchi_book_1d}.

To compute the correlations we absorb the factor $K$ in the
Hamiltonian by rescaling the fields (this preserves the
commutation relation)
\begin{equation}
 \phi = \sqrt{K} \quad,\quad
 \theta = \frac{1}{\sqrt{K}} \tilde\theta
\end{equation}
The fields $\tilde\phi$ and $\tilde\theta$ can be expressed in
terms of bosons operator
$[b^{\phantom\dagger}_q,b^\dagger_{q'}]=\delta_{q,q'}$. This
ensures that their canonical commutation relations are
satisfied. One has
\begin{equation} \label{eq:phitetbos}
\begin{split}
 \phi(x) &=  - \frac{i\pi}L
 \sum_{p\ne 0} \left(\frac{L |p|}{2\pi}\right)^{1/2}
 \frac1p e^{-\alpha|p|/2-i p x}(b^\dagger_p + b^{\phantom\dagger}_{-p}) \\
 \theta(x) &=  \frac{i\pi}L
 \sum_{p\ne 0} \left(\frac{L |p|}{2\pi}\right)^{1/2}
 \frac1{|p|} e^{-\alpha|p|/2-i p x}(b^\dagger_p - b^{\phantom\dagger}_{-p})
\end{split}
\end{equation}
where $L$ is the size of the system and $\alpha$ a short
distance cutoff (of the order of the interparticle distance)
needed to regularize the theory at short scales. The above
expressions are in fact slightly simplified and zero modes
should also be incorporated \cite{giamarchi_book_1d}. This will
not affect the remaining of this section and the calculation of
the correlation functions.

It is easy to check by a direct substitution of
(\ref{eq:phitetbos}) in (\ref{eq:luthamphen}) that Hamiltonian
(\ref{eq:luthamphen}) with $K=1$ is simply
\begin{equation} \label{eq:hamsimlut}
 \tilde H = \sum_{p \ne 0} u |p| b^\dagger_p b^{\phantom\dagger}_p
\end{equation}
The time (or imaginary time \cite{mahan_book}) dependence of
the field can now be easily computed from (\ref{eq:hamsimlut})
and (\ref{eq:phitetbos}). This gives
\begin{equation} \label{eq:timdepphi}
 \phi(x,\tau) =  - \frac{i\pi}L
 \sum_{p\ne 0} \left(\frac{L |p|}{2\pi}\right)^{1/2}
 \frac1p e^{-\alpha|p|/2-i p x}(b^\dagger_p e^{u|p|\tau} + b^{\phantom\dagger}_{-p} e^{-u|p|\tau})
\end{equation}
and a similar expression for $\theta$. In order to compute
physical observable we need to get correlations of exponentials
of the fields $\phi$ and $\theta$. To do so one simply uses
that for an operator $A$ that is \emph{linear} in terms of
boson fields and a quadratic Hamiltonian one has
\begin{equation}
 \langle T_\tau e^{A} \rangle = e^{\frac{1}{2} \langle T_\tau A^2 \rangle}
\end{equation}
where $T_\tau$ is the time ordering operator.  Thus, for
example
\begin{equation}
 \langle T_\tau e^{i2\phi(x,\tau)}e^{-i2\phi(0,0)}\rangle =
 e^{-2\langle T_\tau [\phi(x,\tau)-\phi(0,0)]^2\rangle}
\end{equation}

\begin{figure}
\begin{center}
 \includegraphics[width=0.8\linewidth]{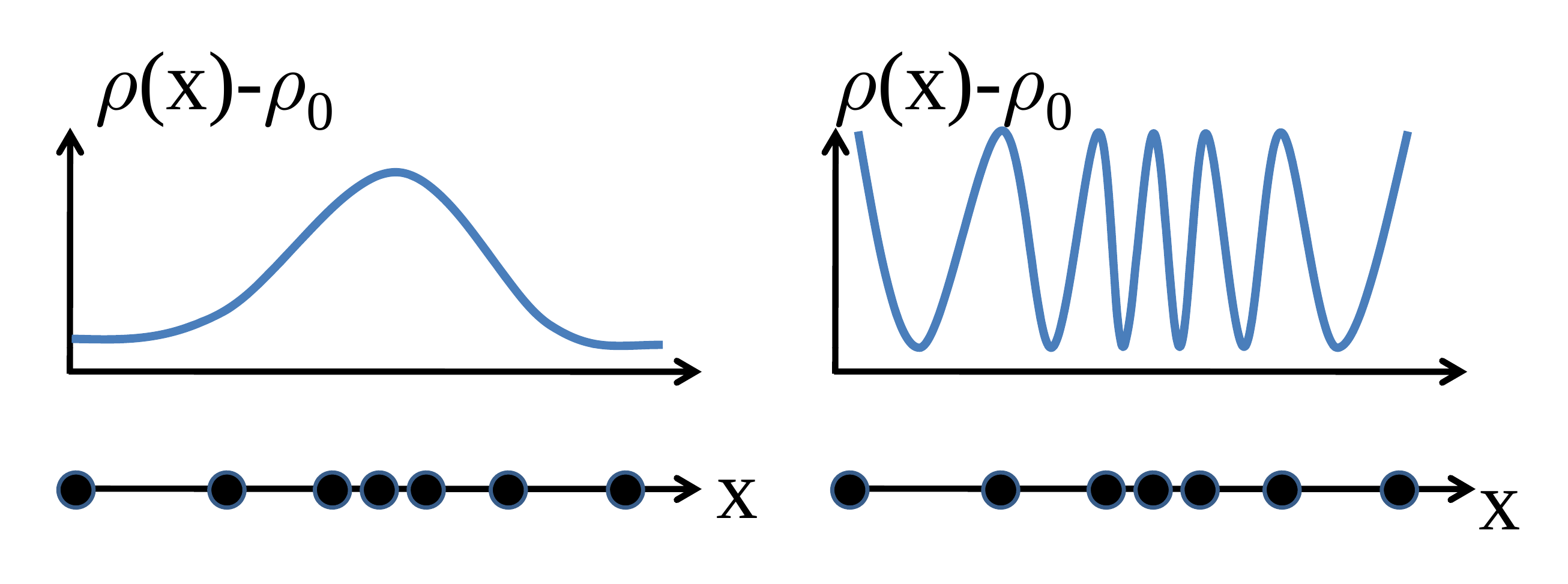}
\end{center}
 \caption{The density $\rho(x)$ can be decomposed in components varying with different Fourier
 wavevectors. The characteristic scale to separate these modes is the inter particle
 distance. Only the two lowest harmonics are represented here.
 Although they have very different spatial variations both these
 modes depends on the \emph{same} smooth field $\phi(x)$.
 (left) the smooth variations of the density at lengthscale larger than the lattice spacing. These are simply
 $-\nabla\phi(x)/\pi$. (right) The density wave corresponding to oscillations of the density at a wavevector $Q=2\pi\rho_0$. These modes
 correspond to the operator $e^{i\pm2\phi(x)}$.}
 \label{fig:densmodes}
\end{figure}

Using these rules it is easy to compute the correlations
\cite{giamarchi_book_1d}. If we want to compute the
fluctuations of the density
\begin{equation}
\langle T_\tau \rho(x,\tau) \rho(0) \rangle
\end{equation}
we obtain, for bosons or fermions, using (\ref{eq:locintbos})
\begin{multline}\label{eq:densluttinger}
 \langle T_\tau \rho(x,\tau) \rho(0) \rangle = \rho_0^2
 + \frac{K}{2\pi^2}
 \frac{y_\alpha^2-x^2}{(x^2+y_\alpha^2)^2}
 + \rho_0^2 A_2 \cos(2\pi\rho_0 x) \left(\frac{\alpha}{r}\right)^{2K}
 \\
 + \rho_0^2 A_4 \;\cos(4\pi\rho_0 x) \left(\frac{\alpha}{r}\right)^{8K} + \cdots
\end{multline}
where $r = \sqrt{x^2 + y^2}$ and $y = u \tau$. Here, the lowest
distance in the theory is $\alpha \sim \rho_0^{-1}$. The
amplitudes $A_i$ are non-universal objects. They depend on the
precise microscopic model, and even on the parameters of the
model. Contrary to the amplitudes $A_n$, which depend on the
precise microscopic model, the power-law decay of the various
terms are \emph{universal}. They \emph{all} depend on the
unique Luttinger coefficient $K$. Physically the interpretation
of the above formula is that the density of particles has
fluctuations that can be sorted compared to the average
distance between particles $\alpha \sim d = \rho_0^{-1}$. This
is shown on \fref{fig:densmodes}.
The fluctuations of long wavelength decay with a universal
power law. These fluctuations correspond to the hydrodynamic
modes of the interacting quantum fluid. The fact that their
fluctuation decay very slowly is the signature that there are
massless modes present. This corresponds to the sound waves of
density described by (\ref{eq:luthamphen}). However the density
of particles has also higher fourier harmonics. The
corresponding fluctuations also decay very slowly but this time
with a non-universal exponent that is controlled by the LL
parameter $K$. This is also the signature of the presence of a
continuum of gapless modes, that exists for Fourier components
around $Q = 2n\pi \rho_0$ as shown in \fref{fig:densmodes}. For
bosons $K$ goes to infinity when the interaction goes to zero
which means that the correlations in the density decays
increasingly faster with smaller interactions. This is
consistent with the idea that the system becoming more and more
superfluid smears more and more its density fluctuations. For
fermions the noninteracting point corresponds to $K=1$ and one
recovers the universal $1/r^2$ decay of the Friedel
oscillations in a free electron gas. For repulsive interactions
$K<1$ and density correlations decay more slowly, while for
attractive interactions $K>1$ they will decay faster, being
smeared by the superconducting pairing.

\begin{figure}
\begin{center}
  \includegraphics[width=0.7\linewidth]{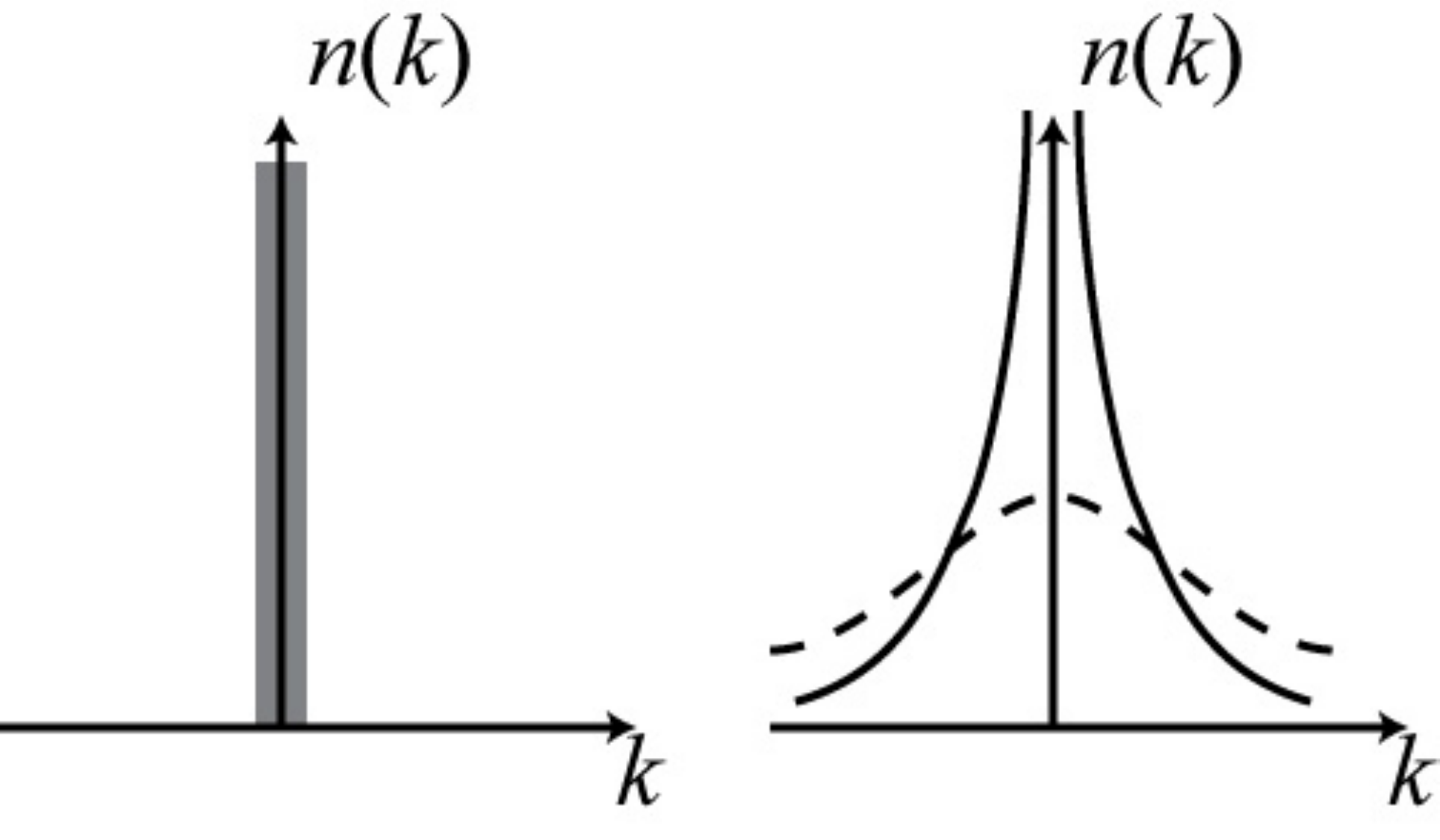}
\end{center}
 \caption{Momentum distribution $n(k)$ for the bosons as a function of the momentum $k$.
 (left) For non-interacting bosons all bosons are in $k=0$ state, thus $n(k)\propto \delta(k)$ (thick line).
 (right) As soon as interactions are introduced a true condensate
 cannot exist. The $\delta$ function is replaced by a power law
 divergence with an exponent $\nu = 1-1/(2K)$ (solid line). At finite temperature,
 the superfluid correlation functions decay exponentially leading to a rounding of the divergence and a lorentzian like
 shape for $n(k)$. This is indicated by the dashed line.}
 \label{fig:singlebos}
\end{figure}

Let us now turn to the single particle correlation function
\begin{equation}
 G(x,\tau) = - \langle T_\tau \psi(x,\tau) \psi^\dagger(0,0) \rangle
\end{equation}
For bosons, at equal time this correlation function is a direct
measure on whether a true condensate exists in the system. Its
Fourier transform is the occupation factor $n(k)$. In presence
of a true condensate, this correlation function tends to
$G(x\to \infty,\tau =0) \to |\psi_0|^2$ the square of the order
parameter $\psi_0 = \langle \psi(x,\tau) \rangle$ when there is
superfluidity. Its Fourier transform is a delta function at
$q=0$, as shown in \fref{fig:singlebos}. In one dimension, no
condensate can exist since it is impossible to break a
continuous symmetry even at zero temperature, so this
correlation must always go to zero for large space or time
separation. Using (\ref{eq:singlebos}) the correlation function
can easily be computed. Keeping only the most relevant term
($p=0$) leads to (I have also put back the density result for
comparison)
\begin{equation} \label{eq:singleboscor}
\begin{split}
  \langle T_\tau \psi(r) \psi^\dagger(0)\rangle &= A_1
  \left(\frac{\alpha}{r}\right)^{\frac1{2K}} + \cdots \\
  \langle T_\tau \rho(r)\rho(0)\rangle &= \rho_0^2 + \frac{K}{2\pi^2}
  \frac{y_\alpha^2 - x^2}{(y_\alpha^2 + x^2)^2} +
  A_3 \cos(2\pi\rho_0 x) \left(\frac{1}{r}\right)^{2K} + \cdots
\end{split}
\end{equation}
where the $A_i$ are the non-universal amplitudes. For the
non-interacting system $K=\infty$ and we recover that the
system possesses off-diagonal long-range order since the
single-particle Green's function does not decay with distance.
The system has condensed in the $q=0$ state. As the repulsion
increases ($K$ decreases), the correlation function decays
faster and the system has less and less tendency towards
superconductivity. The occupation factor $n(k)$ has thus no
delta function divergence but a power law one, as shown in
\fref{fig:singlebos}.
Note that the presence of the condensate or not is not directly
linked to the question of superfluidity. The fact that the
system is a Luttinger liquid with a finite velocity $u$,
implies that in one dimension an interacting boson system has
always a linear spectrum $\omega= u k$, contrary to a free
boson system where $\omega \propto k^2$. Such a system is thus
a \emph{true} superfluid at $T=0$ since superfluidity is the
consequence of the linear spectrum \cite{mikeska_supra_1d}. Of
course when the interaction tends to zero $u\to 0$ as it should
to give back the quadratic dispersion of free bosons.

\begin{figure}
\begin{center}
 \includegraphics[width=0.7\linewidth]{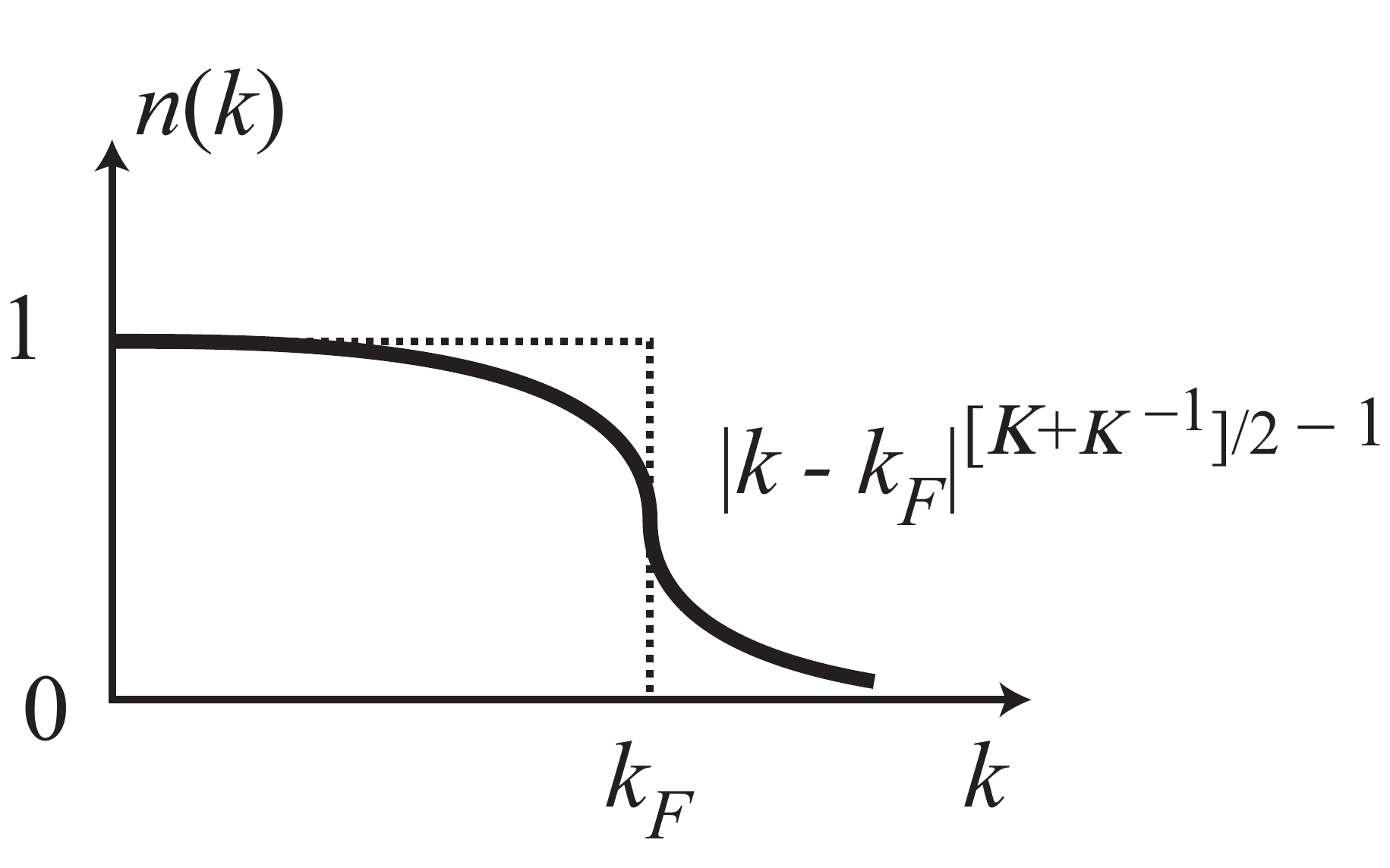}
\end{center}
 \caption{\label{fig:occuplut} The occupation factor $n(k)$. Instead of the usual
 discontinuity at $\kF$ for a Fermi liquid, it has a power law
 essential singularity. This is the signature that fermionic
 quasiparticles do not exist in one dimension. Note that the position of the
 singularity is still at $\kF$. This is a consequence of Luttinger's theorem stating that the volume of the Fermi surface cannot be changed
 by interactions.}
\end{figure}

For fermions the single particle correlation function contains
the terms $p = \pm 1$, corresponding to fermions close to $\pm
\kF$ respectively. If we compute the correlation for the right
movers we get
\begin{equation}
\begin{split}
 G_R(x,\tau) &= -e^{i \kF x} \langle T_\tau e^{i(\theta(x,\tau)-\phi(x,\tau)} e^{-i(\theta(0,0)-\phi(0,0))} \rangle \\
             &= e^{i \kF x} e^{-[\frac{K+K^{-1}}2 \log(r/\alpha) - i {\rm Arg}(y + i x)]}
\end{split}
\end{equation}
The single particle correlation thus decays as a non-universal
power law whose exponent depends on the Luttinger liquid
parameter. For free particles ($K=1$) one recovers
\begin{equation}
 G_R(r) =  - e^{i\kF x}
 e^{-\log[(y_\alpha - i x)/\alpha]} =
 -ie^{i\kF x} \frac1{x+i(v_F \tau
 +\alpha\;{\rm Sign}(\tau))}
\end{equation}
which is the normal function for ballistic particles with
velocity $u$. For interacting systems $K\neq 1$ the decay of
the correlation is always faster, which shows that single
particle excitations do not exist in the one dimensional world.
One important consequence is the occupation factor $n(k)$ which
is given by the Fourier transform of the equal time Green's
function
\begin{equation}
 n(k) = \int dx \; e^{-i k x} G_R(x,0^-) =
  - \int dx \;e^{i (\kF - k) x}
 \left(\frac{\alpha}{\sqrt{x^2+\alpha^2}}\right)^{\frac{K+K^{-1}}2}
 e^{i\;{\rm Arg}(-\alpha + i x)}
\end{equation}
The integral can be easily determined by simple dimensional
analysis. It is the Fourier transform of a power law and thus
\begin{equation}
 n(k) \propto |k-\kF|^{\frac{K+K^{-1}}2-1}
\end{equation}
The occupation factor is shown in \fref{fig:occuplut}.
Instead of the discontinuity at $\kF$ that signals in a Fermi
liquid that fermionic quasiparticles are sharp excitations, one
thus finds in one dimension an essential power law singularity.
Formally, this corresponds to $Z=0$, another signature that all
excitations are converted to collective excitations and that
new physics emerges compared to the Fermi liquid case.

\section{Conclusions}

This concludes this brief tour of interacting quantum fluids.
Of course we have just scratched the surface in these notes.
The problem is an active research field, with extensive efforts
both on the theoretical and the experimental point of view.

The Fermi liquid concept is still one of the cornerstone of our
understanding of such systems, and is certainly the reference
on which all novel properties must be compared. Models allowing
to deal with interactions are still resisting our best attempts
to fully solve them and in particular the Hubbard model, after
about 50 years still remains a challenge, in particular in two
dimensions. Cold atomic systems in optical lattices have
provided a remarkable realization of such models and it is
certain that the stimulation of those novel experimental
realization should help driving the field forward.

Last but not least the one dimensional world, with its own
properties and challenges is now at a stage where it has the
corresponding concept of the Fermi liquid, the so called
Luttinger liquid. This allows to move forward and understand
the effects of various perturbations, such as lattice,
disorder, effect of coupling of one dimensional chains, which
are at the heart of the several realization of the one
dimensional systems that we have today. We thus see that the
one dimensional world can be connected by that route to its
higher dimensional counterpart, and might also provide a way to
tackle it. Considerable development can thus be expected in the
coming years.

\section{Acknowledgements}

The research described in this work was supported in part by
the Swiss NSF under MaNEP and Division II.


\end{document}